\documentclass[reprint,superscriptaddress,amsmath,amssymb,aps,prc]{revtex4-1}

\usepackage{graphicx}
\usepackage{hyperref}
\usepackage{color}
\usepackage[english]{babel}
\usepackage{bm}% bold math
\usepackage[T1]{fontenc}
\usepackage{threeparttable}
\usepackage{multirow}
\usepackage{wrapfig}

\makeatletter
\def\maketitle{
\@author@finish
\title@column\titleblock@produce
\suppressfloats[t]}
\makeatother

\begin{document}

\title{Binding energies, charge radii, spins and moments: odd-odd Ag isotopes and discovery of a new isomer}
\author{B.~van den Borne}
\email{bram.vandenborne@kuleuven.be}
\affiliation{KU Leuven, Instituut voor Kern- en Stralingsfysica, Celestijnenlaan 200D, B-3001 Leuven, Belgium}
\author{M.~Stryjczyk}
\affiliation{University of Jyvaskyla, Department of Physics, Accelerator laboratory, P.O. Box 35(YFL) FI-40014 University of Jyvaskyla, Finland}
\author{R.~P.~de Groote}
\affiliation{KU Leuven, Instituut voor Kern- en Stralingsfysica, Celestijnenlaan 200D, B-3001 Leuven, Belgium}
\affiliation{University of Jyvaskyla, Department of Physics, Accelerator laboratory, P.O. Box 35(YFL) FI-40014 University of Jyvaskyla, Finland}
\author{A.~Kankainen}
\affiliation{University of Jyvaskyla, Department of Physics, Accelerator laboratory, P.O. Box 35(YFL) FI-40014 University of Jyvaskyla, Finland}
\author{D.A.~Nesterenko}
\affiliation{University of Jyvaskyla, Department of Physics, Accelerator laboratory, P.O. Box 35(YFL) FI-40014 University of Jyvaskyla, Finland} 
\author{L.~Al~Ayoubi}
\affiliation{University of Jyvaskyla, Department of Physics, Accelerator laboratory, P.O. Box 35(YFL) FI-40014 University of Jyvaskyla, Finland}
\affiliation{Université Paris Saclay, CNRS/IN2P3, IJCLab, 91405 Orsay, France}
\author{P.~Ascher}
\affiliation{Universit\'e de Bordeaux, CNRS, LP2I Bordeaux, UMR 5797, F-33170 Gradignan, France}
% \author{S.~Ayet}
% \affiliation{II. Physikalisches Institut, Justus Liebig Universitat Gießen, 35392 Gießen, Germany}
\author{O.~Beliuskina}
\affiliation{University of Jyvaskyla, Department of Physics, Accelerator laboratory, P.O. Box 35(YFL) FI-40014 University of Jyvaskyla, Finland}
\author{M.L.~Bissell}
\affiliation{Department of Physics and Astronomy, University of Manchester, Manchester M13 9PL, United Kingdom}
\author{J. Bonnard}
\affiliation{Department of Physics, University of York, Heslington, York YO10 5DD, United Kingdom}
\affiliation{Université de Lyon, Institut de Physique des 2 Infinis de Lyon, IN2P3-CNRS-UCBL, 4 rue Enrico Fermi, 69622 Villeurbanne, France}
\author{P.~Campbell}
\affiliation{Department of Physics and Astronomy, University of Manchester, Manchester M13 9PL, United Kingdom}
\author{L.~Canete}
\altaffiliation[Present address: ]{Northeastern University London, Devon House, 58 St Katharine's Way, E1W 1LP, London, United Kingdom}
\affiliation{University of Jyvaskyla, Department of Physics, Accelerator laboratory, P.O. Box 35(YFL) FI-40014 University of Jyvaskyla, Finland}
\author{B.~Cheal}
\affiliation{Department of Physics, University of Liverpool, Liverpool L69 7ZE, United Kingdom}
\author{C.~Delafosse}
\altaffiliation[Present address: ]{Universit\'e Paris Saclay, CNRS/IN2P3, IJCLab, 91405 Orsay, France}
\affiliation{University of Jyvaskyla, Department of Physics, Accelerator laboratory, P.O. Box 35(YFL) FI-40014 University of Jyvaskyla, Finland}
\author{A.~de Roubin}
\altaffiliation[Present address: ]{Université de Caen Normandie, ENSICAEN, CNRS/IN2P3, LPC Caen UMR6534, F-14000 Caen, France}
\affiliation{University of Jyvaskyla, Department of Physics, Accelerator laboratory, P.O. Box 35(YFL) FI-40014 University of Jyvaskyla, Finland}
\affiliation{Universit\'e de Bordeaux, CNRS, LP2I Bordeaux, UMR 5797, F-33170 Gradignan, France}
\author{C.~S.~Devlin}
\affiliation{Department of Physics, University of Liverpool, Liverpool L69 7ZE, United Kingdom}
\author{T.~Eronen}
\affiliation{University of Jyvaskyla, Department of Physics, Accelerator laboratory, P.O. Box 35(YFL) FI-40014 University of Jyvaskyla, Finland}
\author{R.F.~Garcia Ruiz}
\affiliation{Massachusetts Institute of Technology, Cambridge, Massachusetts 02139, USA}
\author{S.~Geldhof}
\altaffiliation[Present address: ]{GANIL, CEA/DRF-CNRS/IN2P3, B.P. 55027, 14076 Caen, France}
\affiliation{University of Jyvaskyla, Department of Physics, Accelerator laboratory, P.O. Box 35(YFL) FI-40014 University of Jyvaskyla, Finland}
\author{M.~Gerbaux}
\affiliation{Universit\'e de Bordeaux, CNRS, LP2I Bordeaux, UMR 5797, F-33170 Gradignan, France}
\author{W.~Gins}
\affiliation{University of Jyvaskyla, Department of Physics, Accelerator laboratory, P.O. Box 35(YFL) FI-40014 University of Jyvaskyla, Finland}
\author{S.~Gr\'evy}
\affiliation{Universit\'e de Bordeaux, CNRS, LP2I Bordeaux, UMR 5797, F-33170 Gradignan, France}
\author{M.~Hukkanen}
\affiliation{University of Jyvaskyla, Department of Physics, Accelerator laboratory, P.O. Box 35(YFL) FI-40014 University of Jyvaskyla, Finland}
\affiliation{Universit\'e de Bordeaux, CNRS, LP2I Bordeaux, UMR 5797, F-33170 Gradignan, France}
\author{A.~Husson}
\affiliation{Universit\'e de Bordeaux, CNRS, LP2I Bordeaux, UMR 5797, F-33170 Gradignan, France}
\author{P. Imgram}
\altaffiliation[Present address: ]{KU Leuven, Instituut voor Kern- en Stralingsfysica, B-3001 Leuven, Belgium}
\affiliation{Institut f\"ur Kernphysik, Technische Universit\"at Darmstadt, D-64289 Darmstadt, Germany}
\author{\'A. Koszor\'us}
\altaffiliation[Present address: ]{KU Leuven, Instituut voor Kern- en Stralingsfysica, B-3001 Leuven, Belgium}
\affiliation{Department of Physics, University of Liverpool, Liverpool L69 7ZE, United Kingdom}
\author{R.~Mathieson}
\affiliation{Department of Physics, University of Liverpool, Liverpool L69 7ZE, United Kingdom}
\author{I.D.~Moore}
\affiliation{University of Jyvaskyla, Department of Physics, Accelerator laboratory, P.O. Box 35(YFL) FI-40014 University of Jyvaskyla, Finland}
\author{G.~Neyens}
\affiliation{KU Leuven, Instituut voor Kern- en Stralingsfysica, Celestijnenlaan 200D, B-3001 Leuven, Belgium}
\author{I.~Pohjalainen}
\affiliation{University of Jyvaskyla, Department of Physics, Accelerator laboratory, P.O. Box 35(YFL) FI-40014 University of Jyvaskyla, Finland}
\author{M.~Reponen}
\affiliation{University of Jyvaskyla, Department of Physics, Accelerator laboratory, P.O. Box 35(YFL) FI-40014 University of Jyvaskyla, Finland}
\author{S.~Rinta-Antila}
\affiliation{University of Jyvaskyla, Department of Physics, Accelerator laboratory, P.O. Box 35(YFL) FI-40014 University of Jyvaskyla, Finland}
\author{M.~Vilen}
\affiliation{University of Jyvaskyla, Department of Physics, Accelerator laboratory, P.O. Box 35(YFL) FI-40014 University of Jyvaskyla, Finland}
\author{V.~Virtanen}
\affiliation{University of Jyvaskyla, Department of Physics, Accelerator laboratory, P.O. Box 35(YFL) FI-40014 University of Jyvaskyla, Finland}
\author{A.P.~Weaver}
\altaffiliation[Present address: ]{TRIUMF, 4004 Wesbrook Mall, Vancouver, British Columbia V6T 2T3, Canada}
\affiliation{School of Computing, Engineering and Mathematics, University of Brighton, Brighton BN2 4GJ, United Kingdom}
\author{A.~Zadvornaya}
\altaffiliation[Present address: ]{University of Edinburgh, Edinburgh EH9 3FD, United Kingdom}
\affiliation{University of Jyvaskyla, Department of Physics, Accelerator laboratory, P.O. Box 35(YFL) FI-40014 University of Jyvaskyla, Finland}

\begin{abstract}
We report on the masses and hyperfine structure of ground and isomeric states in $^{114,116,118,120}$Ag isotopes, measured with the phase-imaging ion-cyclotron-resonance technique (PI-ICR) with the JYFLTRAP mass spectrometer and the collinear laser spectroscopy beamline at the Ion Guide Isotope Separator On-Line (IGISOL) facility, Jyv\"askyl\"a, Finland.
We measured the masses and excitation energies, electromagnetic moments, and charge radii, and firmly established the nuclear spins of the long-lived states. A new isomer was discovered in $^{118}$Ag and the half-lives of $^{118}$Ag long-lived states were reevaluated. We unambiguously pinned down the level ordering of all long-lived states, placing the inversion of the $I=0^-$ and $I=4^+$ states at $A=118$ ($N=71$). Lastly, we compared the electromagnetic moments of each state to empirical single-particle moments to identify the dominant configuration where possible.

\end{abstract}

\maketitle

\section{Introduction}

Hyperfine structure data and mass measurements on the ground and long-lived isomeric states provide information on their single-particle and collective nuclear properties \cite{Yang2023,Yamaguchi2021}. The former is sensitive to the electromagnetic properties of the nucleus, i.e. magnetic dipole and electric quadrupole moments, charge radii, and nuclear spins \cite{Campbell2016}. The latter can provide excitation energies and masses, and thus the level ordering of the isotope \cite{Dilling2018}. Together, they give insight into the nuclear configuration while also probing its collective behavior and therefore the degree of configuration mixing. These observables allow identification and investigation of structural changes, e.g. (sub)shell closures or shape changes throughout an isotopic chain \cite{Otsuka2020}. 

The region below tin ($Z=50$) is of high interest as it allows for probing shell evolution below the closed proton shell at $Z=50$ and between the neutron shell closures at $N=50,82$. While radioactive isotopes between palladium ($Z=46$) and tin ($Z=50$) have been studied before via mass measurements \cite{Hager2007,Breitenfeldt2010,Hakala2012,Kankainen2013,Atanasov2015,Babcock2018,Nesterenko2020,Manea2020,Izzo2021,Mougeot2021,Nesterenko2023,Nies2023,Jaries2023,Ge2024,Jaries2024a,Ruotsalainen2024} and laser spectroscopy \cite{Yordanov2013,Ferrer2014,Hammen2018,Gorges2019,Yordanov2020,Reponen2021,Geldhof2022,Vernon2022,Karthein2024,Degroote2024}, data on neutron-rich odd-odd Ag isotopes is absent or incomplete.

The masses of neutron-rich silver isotopes have previously been studied using the time-of-flight ion-cyclotron-resonance (ToF-ICR) technique at ISOLTRAP, CERN \cite{Breitenfeldt2010} but the long-lived states could not be separated due to the limited resolving power of the ToF-ICR method. The phase-imaging ion-cyclotron-resonance (PI-ICR) \cite{Eliseev2013,Eliseev2014,Nesterenko2018} technique has a much higher resolving power enabling the measurements of the long-lived states as has already been demonstrated with the even-A silver isotopes in Ref. \cite{Degroote2024}.

Multiple decay spectroscopy studies have been performed on neutron-rich odd-odd Ag isotopes identifying energies and multipolarities of the internal isomeric transitions \cite{Koponen1989,Batchelder2005,Batchelder2012}. However, they were conducted using beams without isomeric or even isobaric purification. As a result, spin assignments, which are based on log\textit{ft}-values and branching ratios \cite{Janas1993,Wang2003,Batchelder2005,Batchelder2012}, are often tentative, sometimes causing conflicting conclusions between different publications \cite{Koponen1989,Janas1993,Wang2003,Rissanen2007}.

The combination of mass and laser spectroscopy measurements at the same facility with the same production method can remedy this situation by providing a complementary insight into the studied isotope, as these techniques are sensitive to different observables. Furthermore, states and excitation energies can be assigned to their spin-parities unambiguously by comparing the yields between the two experiments as demonstrated earlier in Ref. \cite{Degroote2024}. 

% The use of the PI-ICR technique in a Penning trap enables to resolve the long-lived states as it has a much better resolving power than the previously used ToF-ICR \cite{Eliseev2013,Eliseev2014,Nesterenko2018}.

Recent mass spectrometry and collinear laser spectroscopy results \cite{Degroote2024} have been reported on neutron-rich odd-even $^{113-121}$Ag, unambiguously pinning down the spins and excitation energies and showing the continuation of the $I=7/2^+$ and $I=1/2^-$ spins for these isotopes. Notably, the g-factors of the $I=7/2^+$ states show little variation for varying neutron numbers, indicating a relatively pure three-hole ($\pi$g$^{-3}_{9/2})_{7/2}$ configuration, while for the $I=1/2^-$ states they show a clear decreasing trend indicating a varying degree of configuration mixing between different isotopes.

This work presents mass and collinear laser spectroscopy measurements performed in parallel on $^{114,116,118,120}$Ag. The results are combined with the current decay data to firmly establish spin and parity assignments as well as the level ordering for three long-lived states observed in $^{116,118,120}$Ag. The electromagnetic moments are compared to empirical single-particle moments and a dominant configuration is attributed where possible. We already reported on the mean-squared charge radii of the lowest-spin states of $^{114-120}$Ag in a previous publication \cite{Reponen2021}. Here, we re-investigated these radii resulting in larger uncertainties and we provide the electromagnetic moments, masses and excitation energies, and the nuclear mean-squared charge radii for all long-lived states in this mass range. 

% The In($Z=49$) nucleus was long theorized as a textbook example of single-particle behavior in a nucleus, where an effective g-factor was used to explain the difference between the Schmidt limit and the experimental trend \textbf{[add ref]}. Recent laser spectroscopic work on neutron-rich In in combination with ab initio and Density Functional Theory (DFT) calculations, however, have shown that this is not a sufficient explanation as magic $^{131}$In deviates from the constant g-factor trend and shows a sharp rise towards the Schmidt limit \textbf{add ref to adams nature paper}. 

% Paper roughly as follows. We have info on spins of 114,116,118,120. From what I can tell most papers don't mention three states for 116-120, but we did see three.

% We furthermore have mass data for a few more odd-odd nuclei which we can use to better investigate trends in the level energies.

% Aim of the paper could be to put our results in context with literature, assign spins where we can/rule out literature suggestions, but remain somewhat cautious about what we can't directly conclude from laser data.

\section{Experimental methods}

% Specific to the laser spectroscopy, to be expanded with mass measurements
Radioactive silver isotopes were produced at the IGISOL facility by impinging a 25\,MeV proton beam onto a thin uranium foil, where the protons induced the uranium nuclei to fission. Reaction products were stopped in a helium-filled gas cell, operated at 300\,mbar of gas pressure. The gas flow guided these ions into a sextupole ion guide \cite{Karvonen2008}, after which they were accelerated to 30\,kV. The ions were then mass-separated using a 55-degree dipole magnet and injected into a helium-filled radiofrequency quadrupole (RFQ) \cite{Nieminen2001}. Here, the ions were buffer-gas cooled, bunched, and transported to either the JYFLTRAP double Penning trap mass spectrometer \cite{Eronen2012} or the collinear laser spectroscopy beamline \cite{Degroote2020}. The storage time of the ions in the RFQ varied depending on the measurement cycle in the mass measurements and was fixed at 100\,ms for the laser spectroscopy measurements.

At JYFLTRAP, the ions were first cooled and centered on the trap axis by using a mass-selective buffer-gas cooling technique \cite{Savard1991} in the preparation trap, and then transferred to the measurement trap through a 1.5-mm diaphragm. After a few ms, the ions of interest were transferred back to the preparation trap for an additional step of cooling. Two steps were needed for effective cooling and centering of the ions of interest due to the large amount of contaminant ions. For $^{114}$Ag$^+$ ions, a Ramsey-cleaning method \cite{Eronen2008} with an excitation pattern 5-30-5\,ms was additionally applied to remove neighboring isobars. Finally, the ions of interest with charge-to-mass ratio $q/m$ were sent to the measurement trap, where their cyclotron frequency $\nu_c = qB/(2 \pi m)$ in the magnetic field $B$ was determined using the PI-ICR technique \cite{Eliseev2013,Eliseev2014,Nesterenko2018}. 

The phase accumulation time ($t_{acc}$) in the PI-ICR method was chosen in such a way as to separate the isomeric states in the silver isotopes and to ensure that the projection of the cyclotron motion onto the detector does not overlap with any possible isobaric or molecular contamination. The $t_{acc}$ was 400 ms for $^{114,118,120}$Ag and their respective isomeric states while for $^{116}$Ag it was 600 ms. The measurements were performed for singly-charged ions. The magnetic field strength was determined using either $^{133}$Cs (mass excess $\Delta_{lit.} = -88070.943(8)$ keV \cite{Wang2021}) produced in the offline surface ion source \cite{Vilen2020} or the isobaric species produced in fission as the reference. The atomic mass $M$ was determined from the cyclotron frequency ratio $r=\nu_{c,ref}/\nu_{c}$ of the reference ion to the ion of interest as
\begin{equation} \label{eq:mass}
M = (M_{ref} - m_{e}) r + m_{e},
\end{equation}
where $M_{ref}$ and $m_{e}$ are the atomic mass of the reference ion and the electron mass, respectively. The binding energy of the missing electron was neglected.
The systematic uncertainties were included in the final uncertainty of the cyclotron frequency ratios $r$ \cite{Nesterenko2021}. The count-rate class analysis \cite{Roux2013} was performed for the frequency ratios to take into account ion-ion interactions in the measurement trap.

%200 ms for $^{120,122}$Ag and 130 ms for $^{124}$Ag. 

In the laser spectroscopy line, the ions were first neutralized using a charge exchange cell filled with potassium. This cell was heated to about 130 degrees Celsius using a NiCr heating wire wrapped around the center, while the ends were kept at 70 degrees Celsius using a liquid cooling loop filled with GALDEN heat transfer liquid. By changing an acceleration potential applied to the cell, the atoms were Doppler-shifted into resonance with the counter-propagating continuous-wave 328\,nm laser beam to probe the $5s^2\,S_{1/2}\rightarrow5p^2\,P_{3/2}$ transition. The laser wavelength was changed in order to keep the scanning voltage between 0 and 2\,kV. Measurements were spread over three different experimental periods. For the first experimental run, the laser light was produced using an intra-cavity frequency-doubled Spectra Physics 380 dye laser which was pumped by a 5\,W Verdi 532\,nm laser. The other measurements used a Matisse Dye laser pumped by a 10\,W Millennia 532\,nm laser. Instead of intra-cavity doubling, an external WaveTrain frequency-doubler was used. By recording the number of photon counts observed by the photo-multipliers as a function of the wavelength in the rest frame of the atoms, the hyperfine structures of the Ag isotopes could be recorded. To reduce the background from scattered laser light, the recorded number of photons was time-gated such that only the photons measured when the atom bunch was in front of the photo-multiplier were used in the hyperfine spectra. Regular reference measurements were performed on $^{109}$Ag by injecting stable beams produced by a dedicated offline spark-discharge ion source \cite{Vilen2020} into the cooler-buncher. Repeated measurements of the same isotopes were performed across all three experiments as an additional systematic check; all three datasets were found to be in mutual agreement. The hyperfine (HF) spectra were fitted with the SATLAS analysis suite \cite{Gins2018}.

The magnetic dipole moment $\mu$ and spectroscopic electric quadrupole moment $Q$ were determined with Eqs. \ref{EXPEq:dipole} and \ref{EXPEq:quadrupole}.

\begin{eqnarray}
    \label{EXPEq:dipole}
    \mu = \frac{I}{I_{ref}} \frac{A}{A_{ref}} \mu_{ref} \\
    \label{EXPEq:quadrupole}
    Q = \frac{B}{B_{ref}}Q_{ref}
\end{eqnarray}
\\

Here, $I$ is the nuclear spin and $A$ and $B$ are the hyperfine $A$- and $B$-constants. The $\mu_{ref}$ and $Q_{ref}$ denote a non-optical reference measurement of the dipole and quadrupole moment from literature, and the $A_{ref}$ and $B_{ref}$ denote high-precision measurements of the hyperfine $A$- and $B$-constants from literature.

The changes in mean-squared charge radii $\delta\langle r^2 \rangle^{A',A}$ are calculated using Eq. \ref{Eq:chargeR1}.

% The changes in mean-squared charge radii $\langle r^2 \rangle^{A',A}$ are calculated using Eq. \ref{Eq:chargeR1}, which is rewritten as Eq. \ref{Eq:chargeR2} and compared with neighboring chains.

\begin{equation}
    \label{Eq:chargeR1}
    \delta \nu ^{A',A} = FK \delta \langle r^2 \rangle^{A',A} + M \left(\frac{1}{m'}-\frac{1}{m}\right). 
\end{equation}

Here, $\delta\langle \nu \rangle^{A',A}$ is the isotope shift between isotopes \textit{A} and $A'$ which is extracted from the hyperfine spectra, $\delta\langle r^2 \rangle^{A',A}$ is the change in mean-squared charge radius between these isotopes, \textit{m} and $m'$ are their masses, \textit{F} and \textit{M} are the field- and mass-shift atomic factors coming from atomic or empirical calculations and \textit{K} (=0.976) is a correction factor to the field-shift factor for higher-order radial moments as used in Ref. \cite{Reponen2021,Geldhof2022}. As seen in Eq. \ref{Eq:chargeR1}, the $F$ factor is the scaling constant for the contribution due to the difference in nuclear volume to the isotope shift, while the $M$ factor is the scaling constant for the contribution due to the difference in nuclear mass between isotopes \textit{A} and $A'$. 

In the case that the HF structure of multiple nuclear states was overlapped, the highest peak in the HF spectrum was associated with the highest spin nuclear state, as this is the most produced in this experiment \cite{Cannarozzo2023}. Moreover, as the number of peaks in the HF structure is larger than the number of free variables (A$_{\text{lower}}$, A$_{\text{upper}}$, B$_{\text{upper}}$ and isotope shift $\delta\nu$) in $^{116,118,120}$Ag, there is no ambiguity in assigning resonances to the difference nuclear states.

In this work, the statistical uncertainties originate from the uncertainty from the fit and the scan-to-scan variation of each fit parameter. The isotope shifts from the lowest spin states, reported in Ref. \cite{Reponen2021}, are reported here again, including the scan-to-scan variation. This resulted in consistent values, but larger uncertainties. The systematic uncertainties on the nuclear observables originate from the reference moments and hyperfine $A$- or $B$-constants for the dipole and quadrupole moments, and the field- and mass-shift factor and nuclear masses for the changes in mean-squared charge radii. While the hyperfine anomaly in silver can go up to a few percent \cite{Persson2013}, we decided to neglect this contribution as it is not known for the studied isotopes and requires further development. The dipole moment $\mu$ is calculated with a reference  $A_{S_{1/2},109}$ = $-1976.932075(17)$\,MHz \cite{Dahmen1967}, $\mu_{109}$ = $-0.1306906(2)$\,$\mu_N$ \cite{Sahm1974,Mertzimekis2016} assuming a zero hyperfine anomaly, and the quadrupole moment with $B_{P_{3/2},110m}$ = $425(18)$\,MHz \cite{Fischer1975} and $Q_{110m}$ = $1.44(10)$\,b \cite{Berkes1984}. The changes in mean-squared charge radii are extracted with newly-calculated atomic field- and mass-shift factors, F = $-3557(49)$\,MHz/fm$^2$ and M = $1479(14)$ GHz u \cite{Ohayon2024}, different from the ones reported in Ref. \cite{Reponen2021} based on empirical estimates. 

\section{Results and Discussion}

\begin{figure*}[btp]
    \centering
    \includegraphics[width = \textwidth]{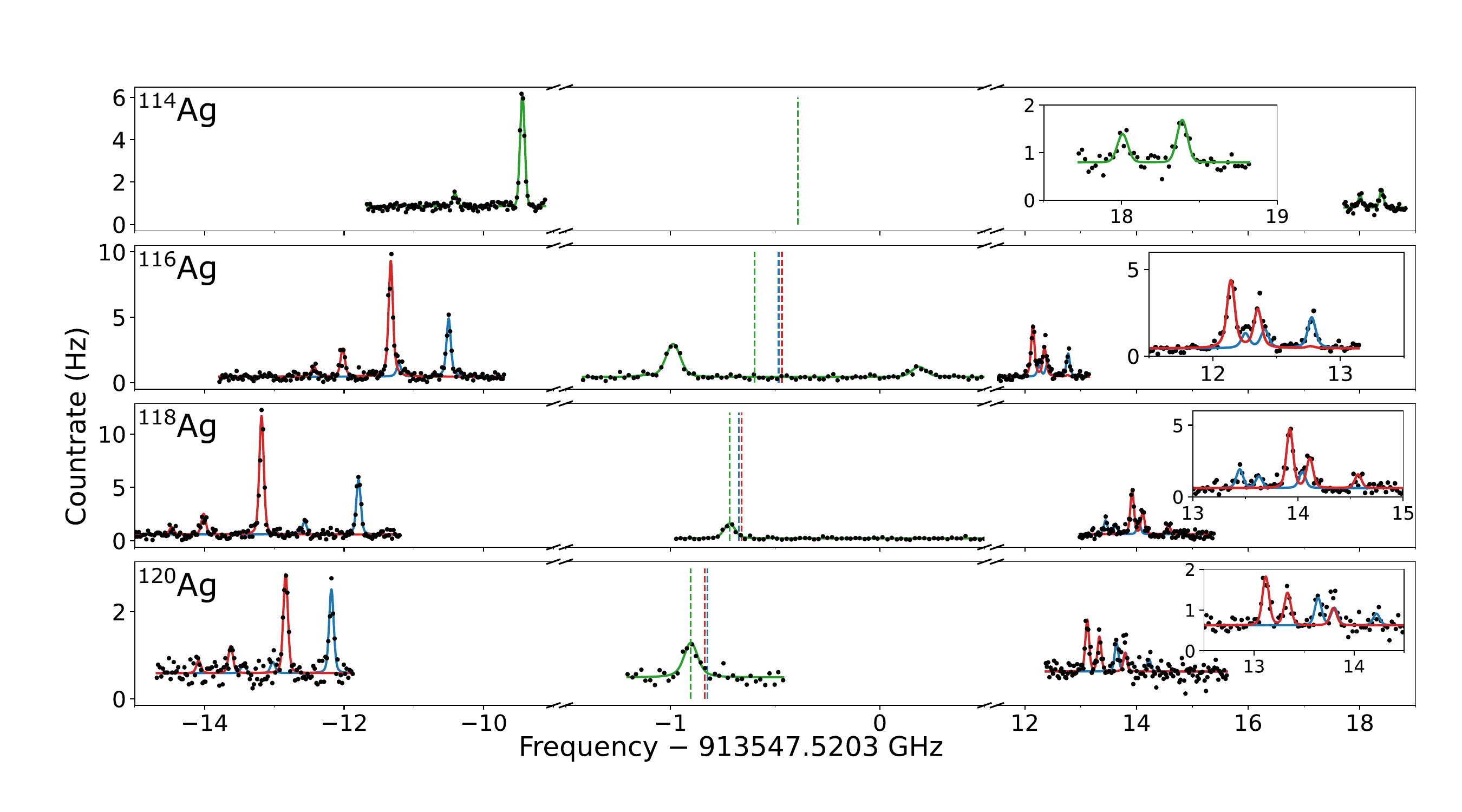}
    \caption{Example hyperfine spectrum of $^{114,116,118,120}$Ag with fitted low-, medium- and high-spin structures (in green, blue, and red respectively), with insets on the high-frequency side. The centroid values are indicated with the dotted line. The differences in background between the separated multiplets are due to the difference in experimental conditions. The x-axis is the measured frequency relative to the transition frequency ($5s^2\,S_{1/2}$ to $5p^2\,P_{3/2}$). Note that the scale in the middle is smaller than in the outer parts.}
    \label{fig:comp114-120}
\end{figure*}

All the results are summarized in Tables \ref{tab:results} and \ref{tab:laserresults}. The masses, spins, and state ordering are discussed in Sec. \ref{subsec:lvlorder_Iassign}, the dipole moments and configurations in Sec. \ref{subsec:dipole_config}, and the quadrupole moments and charge radii in Sec. \ref{subsec:quad_charge}.

\subsection{Level ordering and spin assignments}\label{subsec:lvlorder_Iassign}

The following section discusses the studied silver isotopes' spin assignments and level ordering. Often, spin assignments can be made using laser spectroscopy data based on the ratio of $A$- or $B$-constants of the atomic ground and excited states. However, in this case, the $A(P_{3/2})/A(S_{1/2})$ ratio is rather small ($\sim$1/53.5), which when combined with the size of the experimental uncertainties, prevents such assignments. Furthermore, the $S_{1/2}$ atomic ground state is known to exhibit a large hyperfine anomaly \cite{Persson2013}, further complicating these assignments based on this ratio. We opted to determine the spins in this work for each state separately based on trends in the nuclear observables (e.g. changes in mean-squared charge radii), combined with existing decay spectroscopy data. 

% In case of overlap between isomers, they were identified by the intensities of the hyperfine multiplets, where the multiplet with the highest intensity was assigned to the high spin state since it has a greater production yield.

\subsubsection{$^{114}$Ag}

One long-lived nuclear state, the $I=1^+$ ground state is known in $^{114}$Ag. This state is observed in the mass and laser spectroscopy measurement, the latter indicated in Fig. \ref{fig:comp114-120}. We note that the known $I\leq6^+$ isomer \cite{Penttila1990} has a half-life of 1.50(5)\,ms \cite{NUBASE2020} which is much shorter than our measurement cycles thus it was not present in the beam. The mass excess measured in this work differs from the AME2020 value \cite{Wang2021}, which is based on the ISOLTRAP measurement \cite{Breitenfeldt2010}, by 1.6\,$\sigma$ and it is 4 times more precise.

% In \cite{Koponen1989}, $^{114}$Ag was studied via the decay of $^{114}$Pd to $^{114}$Ag, where a spin $I=1^+$ was deduced. This assignment is confirmed in \cite{Rosner1971,Lund1984} via the decay of $^{114}$Ag to $^{114}$Cd. 
The observed hyperfine structure is consistent with the literature assignment of $I=1$ \cite{Koponen1989} as only two resonances could be observed on the high-frequency side. A nuclear spin $I>1$ would involve three resonances with approximately equal intensities. Furthermore, analysis assuming $I=2$ resulted in a large isotope shift ($932$\,MHz), which would in turn result in an unphysical change in mean-squared charge radius. Our data therefore firmly confirms the spin of the ground state of $^{114}$Ag to be $I=1$. 

\begin{table*}[t]
\centering
    \caption{Results of the mass measurement of the nuclides studied in this work, together with their spins and parities $I^{\pi}$ obtained in this work, and half-lives $T_{1/2}$ from literature \cite{NUBASE2020}.
    Columns Ref. and $r=\nu_{c,ref}/\nu_{c}$ show the reference ions and the measured cyclotron frequency ratios, respectively. Corresponding mass-excess values $\Delta$ and excitation energies of the isomers $E^*$ are tabulated and compared to the literature values $\Delta_{lit.}$ and $E^*_{lit.}$ from Refs. \cite{Wang2021,NUBASE2020}. The $\#$ denotes an extrapolated value from literature \cite{NUBASE2020}. The $^{118}$Ag$^x$ represents an unresolved mixture of two isomeric states in $^{118}$Ag, see text for details. The state ordering of $^{118}$Ag is based on this work while the re-evaluation of the $^{118}$Ag half-lives is explained in detail in section \ref{subsec:118Ag}.}
    \label{tab:results}
    \begin{threeparttable}
    \begin{tabular}{lllllllccc}
        \toprule
Nuclide & $I^{\pi}$ & $T_{1/2}$ & Ref. & $r=\nu_{c,ref}/\nu_{c}$ & $\Delta$ (keV) & $\Delta_{lit.}$ (keV) & ($\Delta-\Delta_{lit.}$) (keV) & $E^*$ (keV) & $E^*_{lit.}$ (keV) \\
 % &  &  &  &  & (keV) & (keV) & (keV) & (keV)\\
\hline
$^{114}$Ag 			& $1^+$  	& 4.6(1) s  & $^{133}$Cs   			& 0.857 066 064(10)     & $-84923.4(12)$ 	& $-84930.8(46)$ & $+7(5)$  &              &	\\
$^{116}$Ag 			& $1^-$  	& 3.83(8) m & $^{116}$Ag$^{m2}$ 	& 0.999 998 740(38) 	& $-82550.6(51)$ 	& $-82542.7(33)$ & $-8(6)$  &              &	\\
$^{116}$Ag$^{m1}$ 	& $4^+$  	& 20(1) s   & $^{116}$Ag$^{m2}$     & 0.999 999 198(27)   	& $-82501.1(43)$ 	& $-82494.8(33)$ & $-6(5)$  & 49.5(51)     & 47.9(1)	\\
$^{116}$Ag$^{m2}$ 	& $7^-$  	& 9.3(3) s  & $^{133}$Cs            & 0.872 134 683(25)    	& $-82414.5(31)$ 	& $-82412.9(33)$ & $-2(5)$  & 136.0(41)    & 129.80(22)	\\
$^{118}$Ag       	& $4^+$  	& 1.92(9)$^1$ s & $^{118}$Ag$^{x}$      & 0.999 998 876(11)    	& $-79541.9(30)$ 	& $-79426.2(25)$ & $-116(4)$ &     0       & 127.63(10)		\\

\multirow{2}{*}{$^{118}$Ag$^{x}$}  	& $0^-$ & $\sim$ 5-6$^1$ s & \multirow{2}{*}{$^{133}$Cs} & \multirow{2}{*}{0.887 207 238(22)} & \multirow{2}{*}{$-79418.4(27)$} & \multirow{2}{*}{$-79553.8(25)$} & \multirow{2}{*}{$+135(4)$} & \multirow{2}{*}{123.5(12)}    &  0\\
& $7^-$ & 1.92(9)$^1$ s & &  &  &  &  &   -- \\

$^{120}$Ag       	& $4^+$  	& 1.52(7) s & $^{133}$Cs            & 0.902 285 837(22)    	& $-75674.1(27)$ 	& $-75651.5(45)$ & $-23(5)$ & 	           &	\\
$^{120}$Ag$^{m1}$   & $0^-$     & 0.94(10) s& $^{120}$Ag            & 1.000 001 030(31)    	& $-75559.0(44)$ 	& $-75650(50)\#$ & $+91(50)$& 115.1(35)    & 0(50)\#	\\
$^{120}$Ag$^{m2}$   & $7^-$  	& 384(22) ms& $^{120}$Ag            & 1.000 001 817(36)    	& $-75471.1(49)$ 	& $-75448.5(45)$ & $-23(7)$ & 203.0(40)    & 203.0(2)	\\
        \hline
    \end{tabular}
       \begin{tablenotes}[]
           \item[(1)] This work
       \end{tablenotes}
\end{threeparttable}
\end{table*}

\begin{table*}[t]
\centering
    \caption{Results of the laser spectroscopy of the nuclides studied in this work together with their spins and parities $I^{\pi}$.
    Columns $A$, $B$ and $\delta\nu$ show the hyperfine $A$- and $B$-constants and isotope shifts of the hyperfine spectra. $\mu$, Q and $\delta\langle$r$^2\rangle^{109,A}$ show the magnetic dipole moments, the spectroscopic electric quadrupole moments, and the changes in mean-squared radii, calculated using $^{109}$Ag as a reference. Statistical and systematic errors are given in parentheses and square brackets, respectively. The isotope shifts for the lowest spin state agree with those in \cite{Reponen2021}, but are reported with a larger uncertainty here, following an investigation of the scan-to-scan variation in the fitted centroids which was not performed in \cite{Reponen2021}.}
    \label{tab:laserresults}
    \begin{threeparttable}
    \begin{tabular}{lllllllll}
        \toprule
Nuclide & $I^{\pi}$ & A(S$_{1/2}$) & $\mu$ & A(P$_{3/2}$) & B(P$_{3/2}$) & Q   & $\delta\nu$  & $\delta\langle$r$^2\rangle^{109,A}$  \\
 &  &  (MHz) & ($\mu_N$) & (MHz) & (MHz) & (b) & (MHz) & (fm$^2$) \\
\hline
$^{114}$Ag 	        & $1^+$ & 19215(22) & $+2.541(14)$ & 352(4)    & 59(3)     &  $+0.201(11)[16]$ & $-853(19)$  &  0.404(5)[6]   \\
$^{116}$Ag 	        & $1^-$ & 805(43)   & $+0.107(6)$  & 15.1(8)$^1$   & $-11(38)$   & $-0.037(129)[3]$ & $-1048(10)$ &  0.522(3)[8] \\
$^{116}$Ag$^{m1}$ 	& $4^+$ & 5329(5)   & $+2.819(15)$ & 96.6(10)  & 234(38)	&  $+0.79(13)[6]$   & $-946(29)$  &  0.494(8)[7] \\
$^{116}$Ag$^{m2}$ 	& $7^-$ & 3304.0(5) & $+3.059(16)$ & 60.10(12) & 293(7)	& $+0.99(3)[8]$     & $-932(31)$  &  0.490(9)[7] \\
$^{118}$Ag  	    & $4^+$ & 5911(2)   & $+3.127(16)$ & 106.0(12) & 297(13)   & $+1.00(4)[8]$     & $-1173(20)$ &  0.618(6)[9] \\
$^{118}$Ag$^{m1}$  	& $0^-$ & 0         & 0         & 0         & 0  	    & 0 	         & $-1222(14)$ &  0.632(4)[9]	\\
$^{118}$Ag$^{m2}$  	& $7^-$ & 3812.9(10)& $+3.53(2)$   & 68.8(4)   & 408(6)	& $+1.38(2)[11]$    & $-1152(15)$ &  0.612(4)[9] \\
$^{120}$Ag       	& $4^+$ & 6045(4)   & $+3.20(2)$   & 107.9(16) & 299(17)   & $+1.01(6)[8]$     & $-1278(4)$  &  0.7067(12)[103] \\
$^{120}$Ag$^{m1}$   & $0^-$ & 0         & 0         & 0         & 0	        & 0              & $-1367(6)$  &  0.7317(17)[106] \\
$^{120}$Ag$^{m2}$   & $7^-$ & 3651(3)   & $+3.38(2)$   & 65.5(9)   & 371(7)    & $+1.25(3)[10]$    & $-1298(8)$  &  0.712(2)[10] \\
        \hline
    \end{tabular}
        \begin{tablenotes}[]
            \item[(1)] The A$(S_{1/2})$/A$(P_{3/2})$ was fixed to the ratio of $^{109}$Ag as the $5p\,^2P_{3/2}$ hyperfine structure was unresolved in our data.
        \end{tablenotes}
\end{threeparttable}
\end{table*}

% In section \ref{subsec:moment106-114}, we further discuss its magnetic moment, and our suggestion of a $\left[\left( \pi g_{9/2}^{-3}\right)_{7/2} \otimes \nu d_{5/2}\right]^{1+}$ configuration.

\subsubsection{$^{116}$Ag}

Three long-lived states in $^{116}$Ag are known in literature \cite{NUBASE2020}. Spins $I=(0^-),(3^+)$ and $(6^-)$ were assigned from $\beta$-decay spectroscopy \cite{Batchelder2005}. All three states were observed and resolved in the mass and laser spectroscopy measurements.

The mass-excess value of the $^{116}$Ag ground state determined at JYFLTRAP differs by 2.4\,$\sigma$ from the AME2020 value \cite{Wang2021}, which is based on the ISOLTRAP measurement \cite{Breitenfeldt2010}. The measured excitation energy of $^{116}$Ag$^{m1}$ is in a good agreement with the NUBASE2020 value while the $^{116}$Ag$^{m2}$ excitation energy differs by 2.8\,$\sigma$ \cite{NUBASE2020}. It should be noted that by using the mass excess of $^{116}$Ag$^{m2}$ from this work and the excitation energy from the NUBASE2020 evaluation \cite{NUBASE2020}, the calculated mass-excess value of the $^{116}$Ag ground state ($-82550.6(51)$\,keV) is in an agreement with AME2020 value.

Fig. \ref{fig:comp114-120} shows the hyperfine structures measured for $^{116}$Ag in this work. Two states exhibit rather large hyperfine splitting, while the third is smaller. Crucially, none of them show only one peak, as would be expected for an $I=0^-$ state. We can thus rule out this spin assignment for the ground state. Therefore, knowing that the long-lived states are connected via E3 transitions \cite{Batchelder2005}, we can infer that the $I=(3^+)$ and $(6^-)$ assignments are also incorrect.

While our data does not provide a clear preference for either spin $1^-$ or $2^-$ for the ground state, the decay spectroscopy strongly favors the $I=1^-$ assignment. In the case of $I=2^-$, a substantial feeding via allowed $\beta$-decays to the $3^-$ states in the daughter $^{116}$Cd isotope can be expected, however, it has not been observed in the decay spectroscopy study of $^{116}$Ag \cite{Batchelder2009}. In addition, a $I=2^-$ would imply that the second isomeric state is $I=8^-$ due to the presence of the E3 transitions, but in Refs. \cite{Batchelder2005,Batchelder2009} feeding to neither 8$^+$ nor (9$^-$) states in $^{116}$Cd was observed. This firmly fixes the spins to $1^-, 4^+$, and $7^-$ for the ground and two isomeric states respectively. We note that these assignments resolve the problem of two missing transitions between proposed (8$^-$) and (5$^+$) states, as reported in Refs. \cite{Porquet2003,Kim2017}.

\subsubsection{$^{118}$Ag}\label{subsec:118Ag}

\begin{figure}[!htb]
    \centering
    \includegraphics[width = \columnwidth]{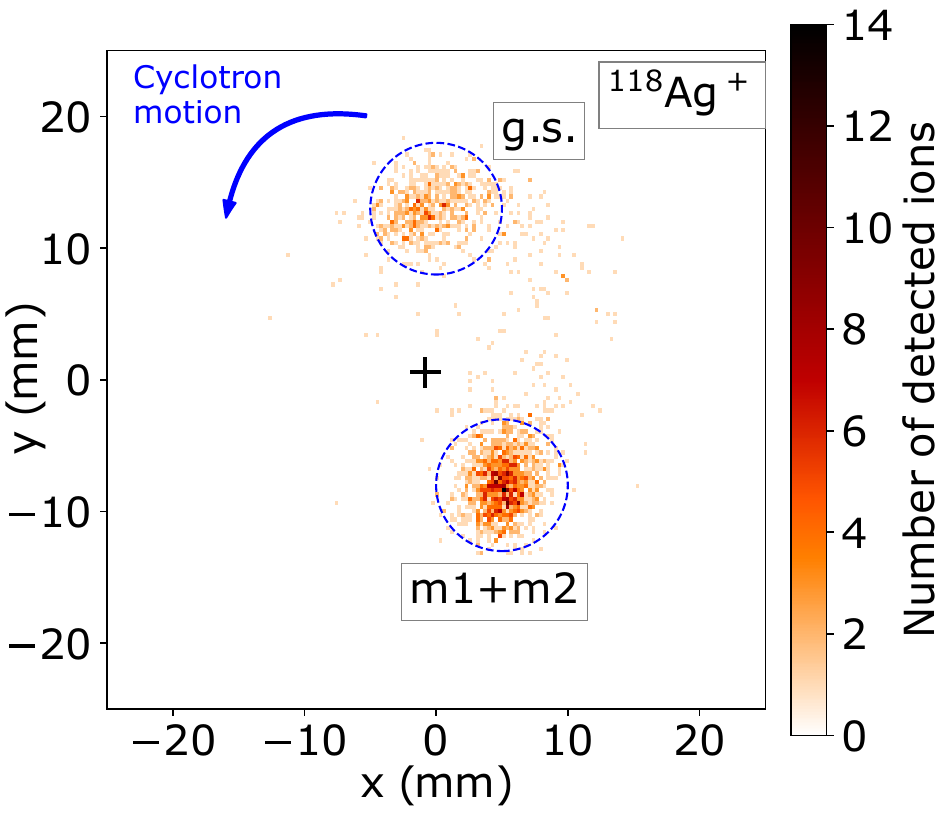}
    \caption{Projection of the cyclotron motion of $^{118}$Ag$^+$ ions onto a position-sensitive detector, obtained with the PI-ICR technique using a phase accumulation time t$_{\text{acc}} = 400$\,ms.}
    \label{fig:PIICR118}
\end{figure}

Two long-lived states in $^{118}$Ag are known in literature \cite{NUBASE2020} and this was also observed by the mass measurement, as shown in Fig. \ref{fig:PIICR118}. However, three states were unambiguously identified in the collinear laser spectroscopy experiment. The phase-accumulation time in the PI-ICR measurement was varied from 50 ms to 800\,ms, and the final measurements were performed with $t_{acc} = 400$\,ms. Most likely the third state, not observed in the PI-ICR measurement, lies close to one of the observed states with a mass difference $\le25$\,keV (based on the accumulation time). The excitation energy of $^{118}$Ag$^{x}$ determined at JYFLTRAP deviates by $-$4.1(12)\,keV from the precisely-known NUBASE2020 value of 127.63(10)\,keV \cite{NUBASE2020}, based on the E3 $\gamma$-decay observed in Refs. \cite{Fogelberg1971,Koponen1989,Rissanen2007}. As this difference is negative, the third state must be lower in energy than 127.63(10)\,keV, thus we conclude that the 127.63(10)\,keV E3 transition connects the ground state and the second isomeric state.

To determine which two states are unresolved in the PI-ICR spectra, we compare the production of each isomer between the mass and laser spectroscopy measurements. In the former, the ion ratio between the ground state and the isomeric spot is about 0.46. The decay losses in the trap were neglected as the phase accumulation time is much shorter than the half-lives of the long-lived states in $^{118}$Ag. In the latter, we calculated the integral of hyperfine peaks of a specific isomer to estimate its relative production rate. Eight different combinations of peaks were considered to verify the consistency of these rates, see Supplementary Material \ref{SupMatSec:production}. The extracted rate of the low-, medium-, and high-spin states is 10.2(4)\%, 30.7(11)\%, and 59.1(13)\%, respectively. These two results are compatible with each other only when the medium-spin state is the ground state and the low- and high-spin states are the excited states, since $30.7(11)/(10.2(4)+59.1(13)) = 0.443(18)$.

\begin{figure}[htb]
    \centering
    \includegraphics[width = \columnwidth]{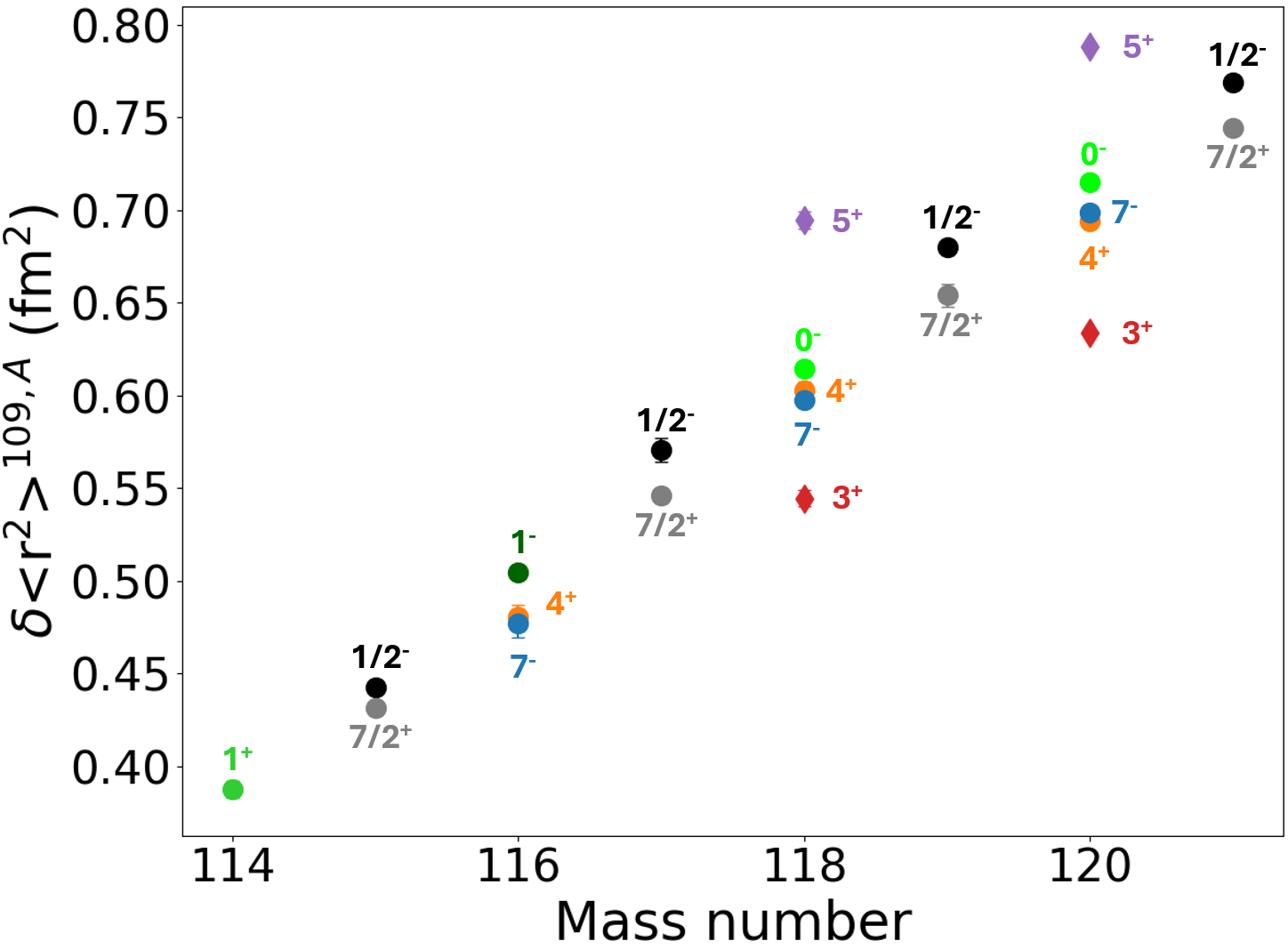}
    \caption{Changes in mean-squared radii for $^{114-121}$Ag with their respective spins. $I=1/2^-$ data reevaluated from Ref. \cite{Reponen2021}, while the $I=7/2^+$ is from this work, see Supplementary Material \ref{SupMatSec:odd-even}. The diamonds show the unphysical charge radii for different spins in $^{118,120}$Ag.}
    \label{fig:chargeradiiAg}
\end{figure}

From the laser spectroscopy data, we can firmly conclude that the low-spin state is a $I=0^-$ state, as only a single peak is observed in the hyperfine structure, shown in Fig. \ref{fig:comp114-120}. The parity is inferred from the only possible configuration which can yield a spin zero in this region of the chart, coupling a proton in the p$_{1/2}$ and a neutron in the s$_{1/2}$ orbital. Additionally, we conclude that the medium-spin state is an $I=4$ state, as a spin 3 or 5 would yield $\langle r^2 \rangle^{109,A}$ of 0.7069(55) or 0.5291(51)\,fm$^2$ respectively, which are very far from the trend shown in Fig. \ref{fig:chargeradiiAg}. Consequently, the E3 transition observed in literature must be between this state and the high-spin state, as an E3 $\gamma$-transition cannot connect the $I=4$ and the $I=0$ states. This fixes the high-spin state to $I=7^-$, where the parity is inferred as this can only be reproduced by coupling a proton in the g$_{9/2}$ and a neutron in the h$_{11/2}$ orbital. Thus, the $I=4$ state must be positive parity.

Based on the combined collinear laser spectroscopy and mass measurements, we firmly assign the ground state as $I=4^+$, the first isomer as $I=0^-$, and the second isomer as $I=7^-$. Using the precisely determined $\gamma$-ray energy from \cite{NUBASE2020}, the $I=7^-$ state is at 127.63(10)\,keV, while the excitation energy of the $I=0^-$ isomer can be estimated to be between 103-123\,keV. The lower limit is determined by the excitation energy of the second isomer and the resolution of the PI-ICR measurement, i.e. $127.63 - 25 \approx 103$\,keV, while the upper limit is the excitation energy of $^{118}$Ag$^x$ from the PI-ICR measurement, 123.5(12)\,keV. Similarly to $^{116}$Ag, the observation of the $I=7^-$ isomeric state allows for the removal of two suggested but unobserved transitions from the excited states scheme in $^{118}$Ag \cite{Wang2017,Kim2017}.

\begin{figure}[htb]
    \centering
    \includegraphics[width = \columnwidth]{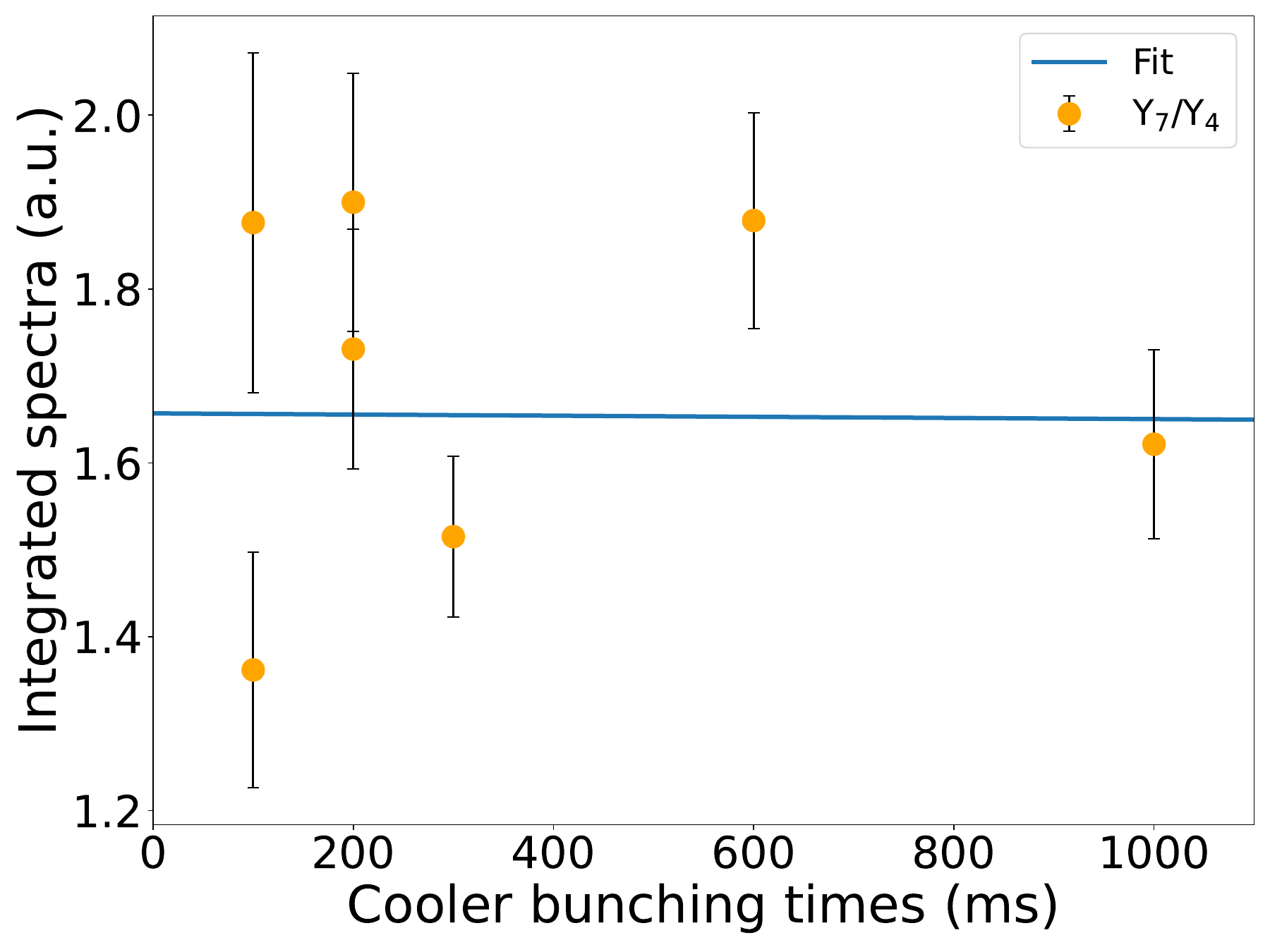}
    \caption{Ratio of integrated hyperfine structure spectra normalised over acquisition time of $I=7^-$ over $I=4^+$ with exponential fit for $^{118}$Ag.}
    \label{fig:RatioHalflife}
\end{figure}

In addition to the hyperfine structure scans of the $I=4^+$ and $I=7^-$ states, multiple collinear laser spectroscopy measurements on these states have been performed with cooler-buncher cooling times, ranging from 100\,ms up to 1\,s. This should, in principle, allow for half-life extraction. However, changes in experimental conditions, such as primary beam current, and transmission losses also influence the number of atoms, so we opted for a relative measurement to cancel out these changes. The ratio of relative yields $Y_7/Y_4$, normalized for acquisition time, is proportional to the activity ratio $A_7/A_4$. By analyzing the change of these ratios as a function of cooling time, as is done in Fig. \ref{fig:RatioHalflife}, one can extract the difference $\Delta\lambda$ between decay constants of the $I=7^-$ ($\lambda_7$) and the $I=4^+$ ($\lambda_4$) state:

\begin{equation}
    \frac{Y_7}{Y_4}(t) = N\times e^{-\Delta\lambda t},
\end{equation}

where $N$ is a normalization constant and $\Delta\lambda = \lambda_7 - \lambda_4$. The fitted difference $\Delta\lambda$ is 4(4)$\times10^{-6}$\,/s, i.e. consistent with the two states having the same half-life within our experimental uncertainties. 

While the existence of the $I=7^-$ state was unknown prior to this work, its half-life can be extracted based on the previous measurements of the 127.63(10)\,keV $\gamma$-ray. Two values have been reported in the literature, 1.9(1)\,s in Ref. \cite{Koponen1989} and 2.0(2)\,s in Ref. \cite{Chrien1978}. We take their weighted average, 1.92(9)\,s as the half-life of the $I=7^-$ state. Based on these results and our measurement, we extracted the half-life of the $I=4^+$ state to be 1.92(9)\,s. The $I=0^-$ state will likely have a half-life of the order of $\sim$5-6\,s as reported in Refs. \cite{Fritzs1967,Weiss1968}, rather than the more recently reported 3.7(2)\,s and 3.76(15)\,s \cite{Chrien1978,Koponen1989} which were measured for isomeric mixtures.

The discovery of a new isomer and the new order of the long-lived states imply a different interpretation of the two decay spectroscopy studies of $^{118}$Ag \cite{Wang2003} and $^{118}$Pd \cite{Janas1993} where only two long-lived states in $^{118}$Ag with different spins and a different level order were proposed. In both of these publications, the production method was the same as in this work and the beam contained the entire isobar, thus, all three long-lived states of $^{118}$Ag were present. As a result, these decay schemes should be reevaluated. In particular, as the half-lives of the $I = 4^+$ and the $I = 7^-$ states determined in this work are identical, the decays of these two states could not be separated in Ref. \cite{Wang2003}. Additionally, in Ref. \cite{Janas1993}, there was no sensitivity to distinguish between the $\beta$ decay of $^{118}$Pd and the internal transition decay of $^{118}$Ag$^{m2}$ as the half-lives of $^{118}$Pd (T$_{1/2}$ = 1.9(1)\,s) and $^{118}$Ag($I = 4, 7$) (T$_{1/2}$ = 1.92(9)\,s) are identical within experimental uncertainties. Consequently, the population of the $I = 7^-$ 127.63(10)\,keV state in the $\beta$-decay of $^{118}$Pd remains doubtful.

\subsubsection{$^{120}$Ag}

% \textcolor{blue}{Marek: I think it should be commented why the high-spin isomer is \textit{not} the strongest spot in pi-icr. I believe it is simply related to the length of the accumulation time/measurement cycle and the difference in half-lives between the isomer and the ground state, but please, correct me if I'm wrong.}

\begin{figure}[htb]
    \centering
    \includegraphics[width = \columnwidth]{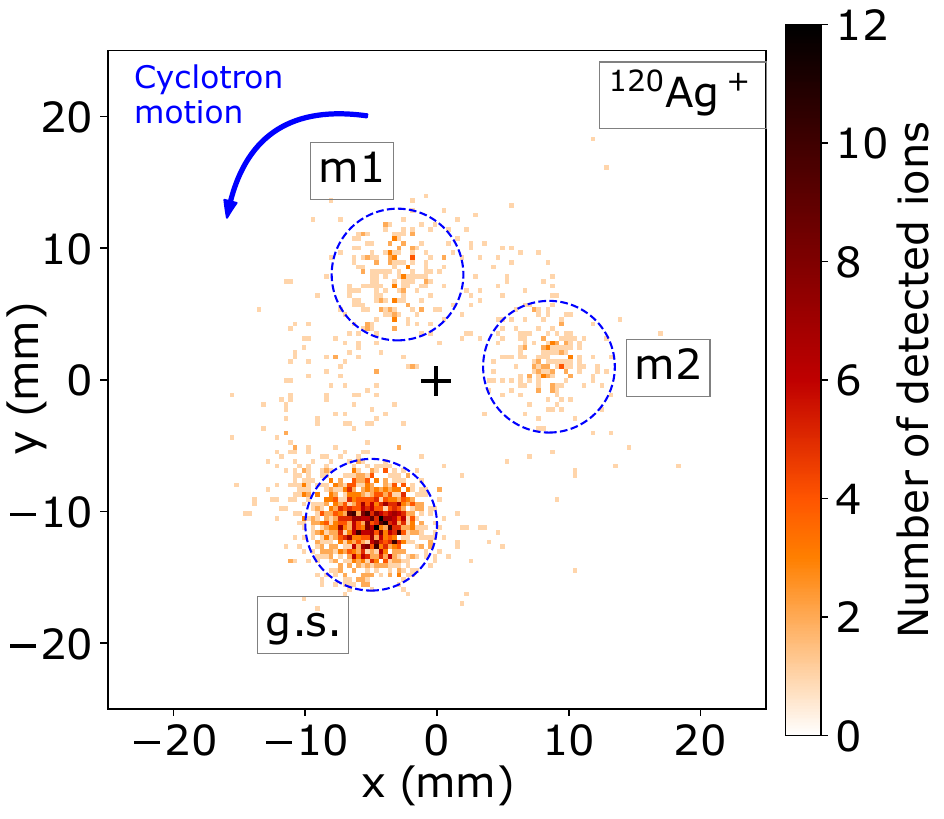}
    \caption{Projection of the cyclotron motion of $^{120}$Ag$^+$ ions onto a position-sensitive detector, obtained with the PI-ICR technique using a phase accumulation time t$_{\text{acc}} = 400$\,ms.}
    \label{fig:PIICR120}
\end{figure}

Three long-lived states in $^{120}$Ag are known in literature \cite{NUBASE2020}, all of which are observed in the mass and laser spectroscopy measurements, see Figs. \ref{fig:comp114-120} and \ref{fig:PIICR120}. Spin-parities $I=(0,1^-), 4(^+)$ and $7(^-)$ were previously assigned based on the $\beta$-feeding to the states in Cd and systematics in Ref. \cite{Batchelder2012}. However, the relative order of the low- and the medium-spin states remained unknown. 

The energy difference between $^{120}$Ag$^{gs}$ and $^{120}$Ag$^{m2}$ measured in this work, 203.0(40)\,keV, matches the known energy difference between the $I=4(^+)$ and $I=7(^-)$ state of 203.0(2)\,keV \cite{NUBASE2020}. Therefore, we conclude that the observed isomer at 115.1(35)\,keV is the low-spin state and its excitation energy was determined for the first time. The mass-excess value of $^{120}$Ag determined at JYFLTRAP differs from the AME2020 value \cite{Wang2021} by 5\,$\sigma$. However, it is exclusively based on the ISOLTRAP value \cite{Breitenfeldt2010} where the three long-lived states were not resolved.

%All three long-lived states that were observed using PI-ICR were also observed in the collinear laser spectroscopy. Spins $I=(0,1^-), 4(^+), 7(^-)$ were assigned based on the $\beta$-feeding to the states in Cd and systematics in \cite{Batchelder2012}. Note however that they were not able to separate the low and medium spin states. 

Our hyperfine spectra feature only a single peak for the low-spin isomer of $^{120}$Ag, as is the case with $^{118}$Ag. This means that there is no hyperfine splitting present which is only possible for a $I=0$ state or a $I>0$ state with a near-zero dipole and quadrupole moment. However, for a $I>0$ state, conversion electrons at 110-115\,keV would be expected in Ref. \cite{Batchelder2012}, but none are seen in their spectra. Thus, we firmly assign a $I=0^-$ to this state, where the parity is inferred as this spin can only be reproduced by coupling a proton in the p$_{1/2}$ and a neutron in the s$_{1/2}$ orbital. 
% Decay spectroscopy measurements in Ref. \cite{Batchelder2012} support this assignment. 
% An $I=1^-$ assignment can be excluded as conversion electrons at 110-115 keV would be expected in Ref. \cite{Batchelder2012}, but no conversion electrons are seen in their spectra.

In Ref. \cite{Batchelder2012}, the absence of $\beta$-feeding to the $I=9^-$ state and the observation of $\beta$-feeding to $I=8^+$ in the daughter $^{120}$Cd led to the $I=7(^-)$ assignment for the high-spin state. This state decays with an E3 $\gamma$-transition to the medium-spin state which is then an $I=4(^+)$ state. The changes in the mean-squared charge radii determined in this work support this assignment. Analysis assuming an $I=3$ or $I=5$ yield $\delta\langle r^2 \rangle^{109,A}$ of 0.7939(12) or 0.6112(10)\,fm$^2$ respectively, which are very far from the trend shown in Fig. \ref{fig:chargeradiiAg}. Therefore we conclude that the $I=4^+$ and $I=7^-$ are the spins for these two states. The parities are assigned with certainty as the $I=7$ can only be reproduced by coupling a proton in the g$_{9/2}$ and a neutron in the h$_{11/2}$ orbital. We note that the ordering in $^{120}$Ag is the same as in $^{118}$Ag thus our measurement firmly places the state order change at $N=71$.

\subsection{Magnetic dipole moments and configuration}\label{subsec:dipole_config}

The following section discusses the magnetic dipole moments and configuration of the different silver isotopes measured in this work. Empirical single-particle moments are calculated with the addition rule from Ref. \cite{Heyde1994}, given by:

\begin{equation}\label{Eq:dipole}
\begin{split}
    \mu = \frac{I}{2}\Bigg[\frac{\mu_{\pi}}{I_{\pi}} + \frac{\mu_{\nu}}{I_{\nu}} + \left(\frac{\mu_{\pi}}{I_{\pi}} - \frac{\mu_{\nu}}{I_{\nu}}\right) \\
    \times \left(\frac{I_\pi(I_\pi +1) - I_\nu(I_\nu +1)}{I(I+1)}\right)\Bigg].
\end{split}
\end{equation}

In this formula, we take the dipole moment of a neighboring odd-$Z$, even-$N$ isotope for $\mu_\pi$, and of a neighboring even-$Z$, odd-$N$ isotope for $\mu_\nu$ to calculate the empirical single-particle dipole moment of the odd-$Z$, odd-$N$ configuration.

\subsubsection{$^{106-114}$Ag}\label{subsec:moment106-114}

\begin{figure}[htb]
    \centering
    \includegraphics[width = \columnwidth]{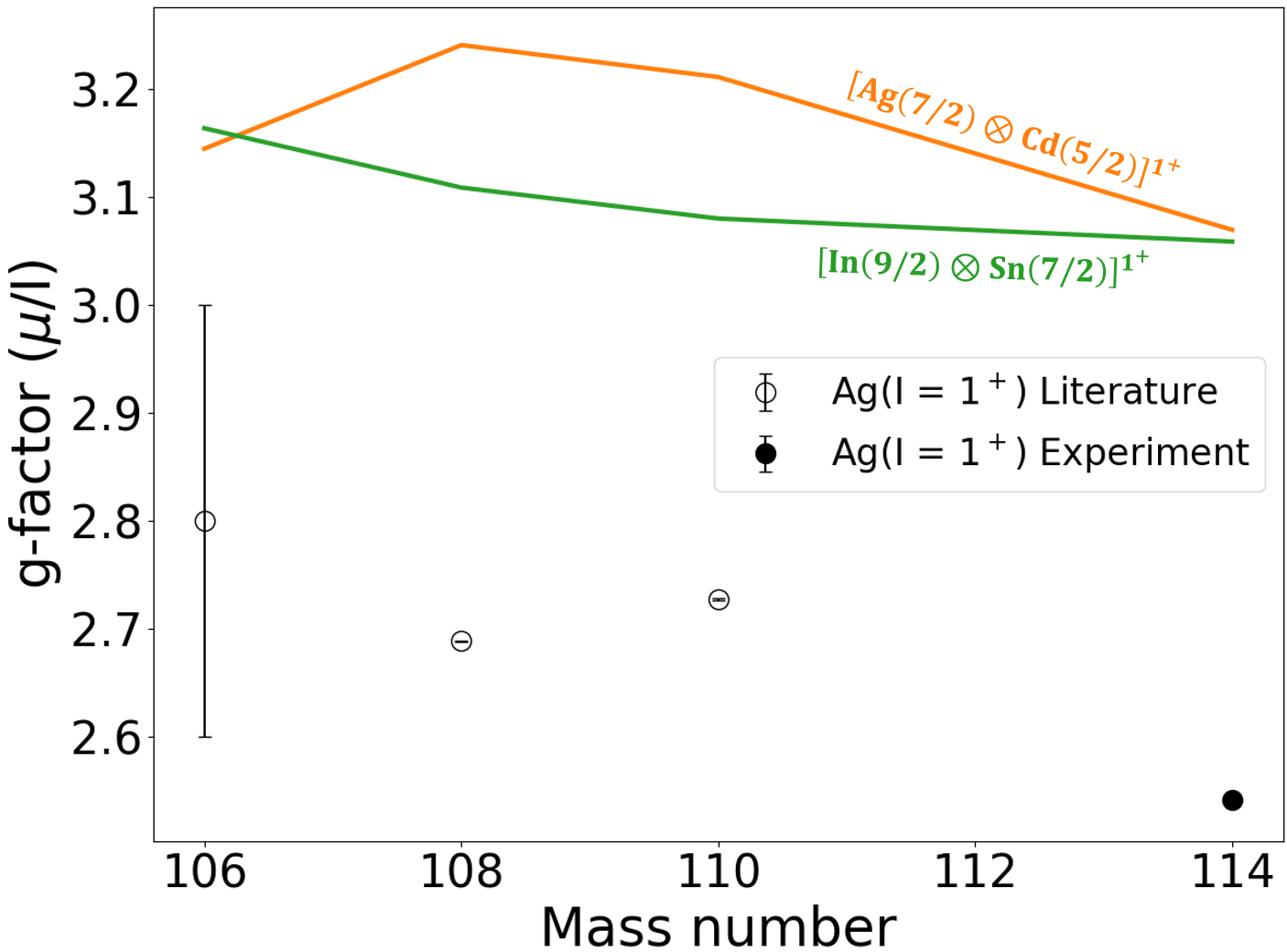}
    \caption{Overview of experimental g-factors of $^{106-114}$Ag($I=1^+$) states with possible empirical single-particle moments shown as solid lines. Open (full) black dots show literature (experimental) results from Refs. \cite{Greenebaum1974,Winnacker1976}. The g-factors used for the empirical single-particle moment calculation are for Ag($I=7/2^+$) from Refs. \cite{Eder1985,Dinger1989,Degroote2024}, for Cd($I=5/2^+$) from Ref. \cite{Spence1972}, for In($I=9/2^+$) from Ref. \cite{Eberz1987} and for Sn($I=7/2^+$) from Ref. \cite{Eberz1987SN}.}
    \label{fig:gfac_spin1}
\end{figure}

The $g$-factors of the $I=1^+$ states follow a relatively constant trend across the isotopic chain, slightly decreasing at $^{114}$Ag as seen in Fig. \ref{fig:gfac_spin1}. The g-factors of the $I=1^+$ states are in reasonable agreement with the calculated moments using $\left[\text{Ag}(7/2)\otimes \text{Cd}(5/2)\right]^{1+}$ and $\left[\text{In}(9/2)\otimes \text{Sn}(7/2)\right]^{1+}$- which can be interpreted as the $\left[\left( \pi g_{9/2}^{-3}\right)_{7/2} \otimes \nu d_{5/2}\right]^{1+}$ and the $\left[ \pi g_{9/2} \otimes \nu g_{7/2}\right]^{1+}$ configuration. While the magnitude and trend of the empirical single-particle moments are similar for both configurations, previous literature suggests a dominating $\left[\left( \pi g_{9/2}^{-3}\right)_{7/2} \otimes \nu d_{5/2}\right]^{1+}$ configuration \cite{Fischer1975,Winnacker1976,Golovko2010}. In Ref. \cite{Ferrer2014} an evolution from a single-proton hole in the $g_{9/2}$ orbital to a three-proton hole in the $g_{9/2}$ orbital from $N=53-59$ is reported for the high-spin states of these isotopes.

\subsubsection{$^{116-120}$Ag}\label{subsec:moment116-120}

The small g-factor of 0.107(6) for the $I=1^-$ state of $^{116}$Ag shows an indication that the $\pi$g$_{9/2}$ and $\nu$h$_{11/2}$ orbitals are unlikely to play a role in this state as its empirical single-particle moment $\big[$In$(\pi g_{9/2})\otimes$Cd$(\nu h_{11/2})\big]^{1^-} = -3.35$ and single-particle moment ($=-4.53$) differs strongly from the experimental g-factor. The best-matching empirical single-particle moment is the $\big[$Ag$(I=1/2)\otimes$Cd$(I=1/2)\big]^{1^-} = -0.82$, which is also not in good agreement with the experimental g-factor. The configurations for which the single-particle moments agree best are $\big[\pi$p$_{1/2}\otimes\nu$(d$_{3/2})_{1/2}\big]^{1^-} = 0.12$ and $\big[\pi$p$_{1/2}\otimes\nu$(g$_{7/2})_{1/2}\big]^{1^-} = -0.05$. We conclude that the configuration of this state is very mixed.

% $\big[$Ag$(\pi$mixed$_{1/2})\otimes$Cd$(\nu$mixed$_{1/2})\big]^{1^-} = -0.82$

% other possible configurations with schmidt moments ($\pi$p$_{1/2}\otimes\nu$s$_{1/2})^{1^-} = -2.18$, ($\pi$p$_{1/2}\otimes\nu$d$_{3/2})^{1^-} = 1.09$,  ($\pi$p$_{1/2}\otimes\nu$(d$_{3/2})_{1/2})^{1^-} = 0.12$, ($\pi$p$_{1/2}\otimes\nu$(g$_{7/2})_{3/2})^{1^-} = 0.66$, ($\pi$p$_{1/2}\otimes\nu$(g$_{7/2})_{1/2})^{1^-} = -0.05$, ($\pi$p$_{1/2}\otimes\nu$(d$_{5/2})_{3/2})^{1^-} = -0.82$, and ($\pi$p$_{1/2}\otimes\nu$(d$_{5/2})_{1/2})^{1^-} = -0.65$. 

\begin{figure}[htb]
    \centering
    \includegraphics[width = \columnwidth]{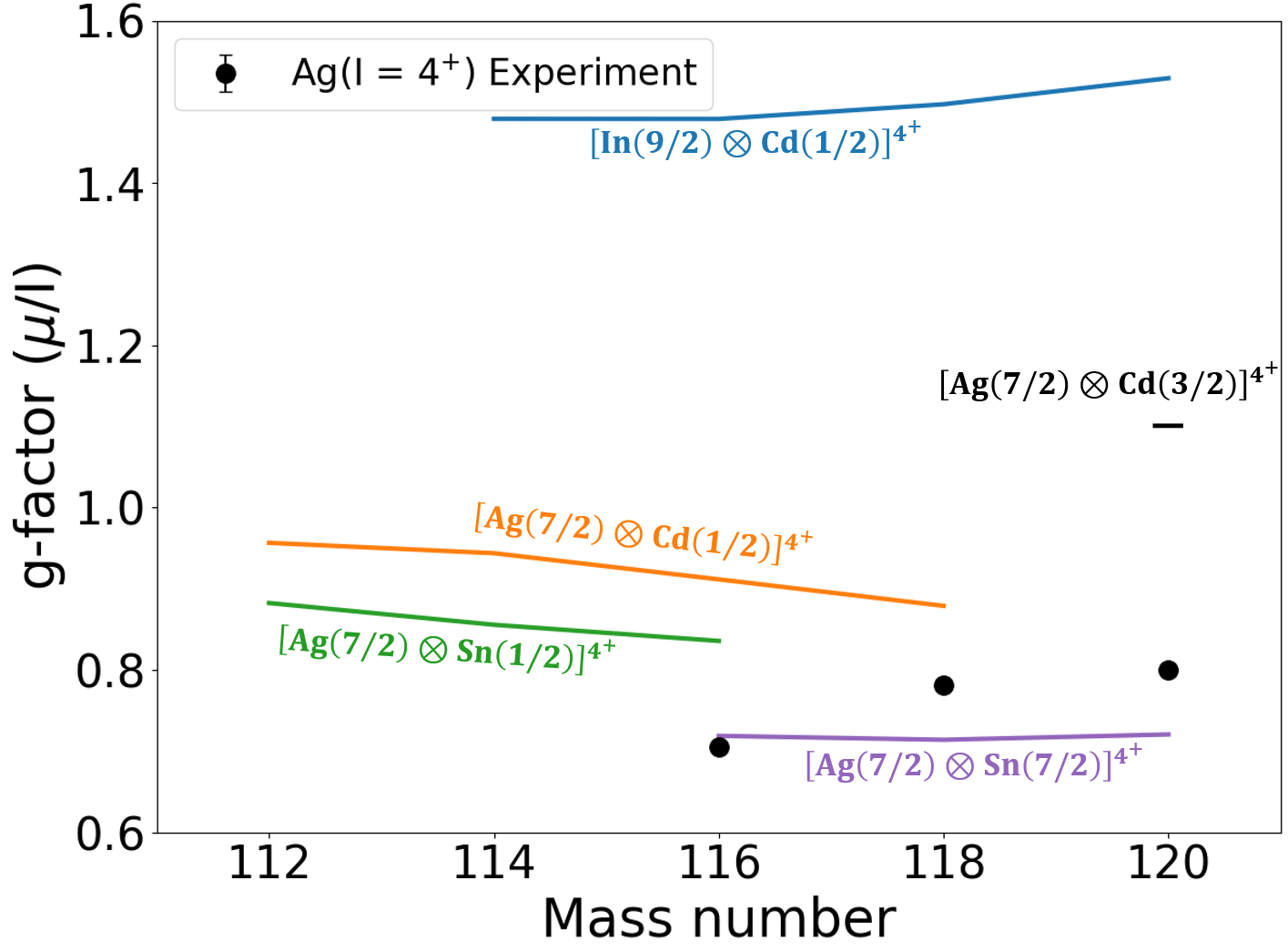}
    \caption{Overview of experimental g-factors of $^{116-120}$Ag$(I=4^+)$ states shown as black dots, with possible empirical single-particle moments shown with solid lines. The g-factors used in the empirical single-particle moment calculation are for Ag($I=7/2$) from Ref. \cite{Degroote2024}, for Cd($I=1/2,3/2$) from Refs. \cite{Chaney1969,Spence1972,Yordanov2013}, for In($I=9/2$) from Ref. \cite{Eberz1987} and for Sn($I=1/2,7/2$) from Ref. \cite{Proctor1950,Eberz1987SN}.}
    \label{fig:gfac_spin4}
\end{figure}

Fig. \ref{fig:gfac_spin4} compares the $I=4^+$ states with the results of the additivity rule using several empirical moments. A proton in the (g$_{9/2}^{-3})_{7/2}$ (Ag(7/2)) configuration coupled with a neutron in the single-particle g$_{7/2}$ (Sn(7/2), a neutron in the mixed Sn$(I=1/2^+)$, or a neutron in the mixed Cd$(I=1/2^+)$ configuration result in similar agreement with the experimental g-factors as seen in Fig. \ref{fig:gfac_spin4}). Likely, there is a high amount of mixing in the neutron configurations, which is expected due to the high density of positive parity orbitals between \textit{N} = 50 and \textit{N} = 82. 

\begin{figure}[htb]
    \centering
    \includegraphics[width = \columnwidth]{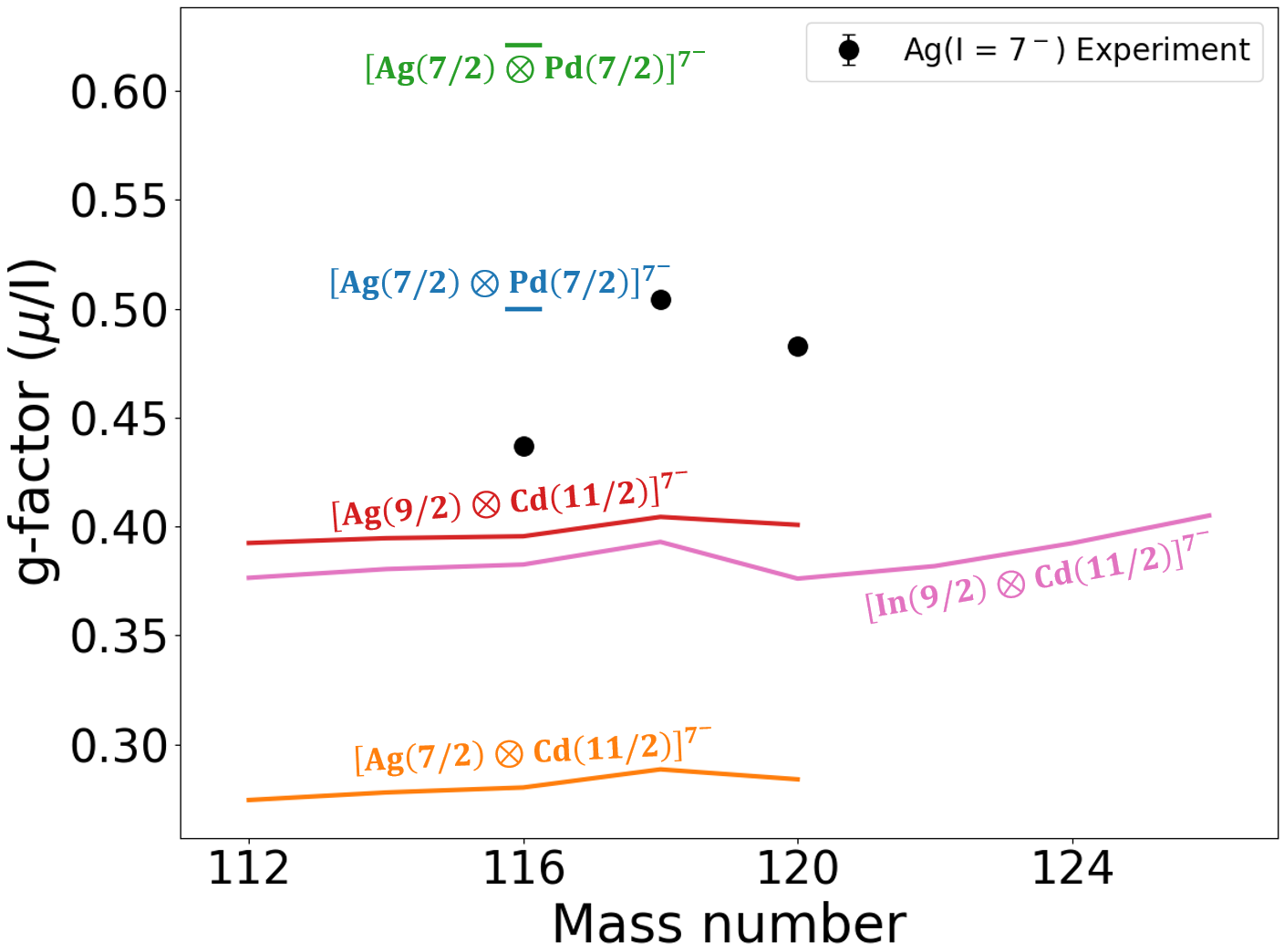}
    \caption{Overview of experimental g-factors of $^{116-120}$Ag$(I=7^-)$ states shown as black circles, with possible empirical single-particle moments shown with solid lines. The g-factors used in the empirical single-particle moment calculations are for Ag($I=7/2$) from Ref. \cite{Degroote2024}, for Cd($I=11/2$) from Ref. \cite{Yordanov2013}, for In($I=9/2$) from Ref. \cite{Eberz1987} and for Pd($I=7/2$) from Ref. \cite{Ortiz2023}.}
    \label{fig:gfac_spin7}
\end{figure}

Lastly, the $I=7^-$ isomers shown in Fig. \ref{fig:gfac_spin7} are in relatively good agreement with the empirical single-particle moments, as expected since fewer orbitals can couple to a $I=7^-$ state and thus less configuration mixing is possible. The closest-matching empirical moments are the $\left[\text{Ag(9/2)}\otimes \text{Cd(11/2)}\right]^{7-}$ and the $\left[\text{Ag(7/2)}\otimes \text{Pd(7/2)}\right]^{7-}$, interpreted as the $\pi$g$_{9/2} \otimes\nu$h$_{11/2}$ and ($\pi$g$_{9/2}^{-3})_{7/2} \otimes$Pd$(I=7/2)$ configurations, both of which yield similar agreement with the data. However, no dominating configuration for either proton or neutron can be deduced from the g-factors.

\subsection{Charge radii and electric quadrupole moments}\label{subsec:quad_charge}

% \begin{figure*}[t]
%     \centering
%     \includegraphics[width = \textwidth]{LevelScheme.PNG}
%     \caption{Reevaluated level scheme for $^{114,116,118,120}$Ag \cite{Penttila1990,Rosner1971,Batchelder2009,Batchelder2012}. The asterisk (*) at the E3 gamma transition at $^{118}$Ag depicts that the branching ratio is not known, as previous publications overestimated the $\beta$-decay of the excited $I=7^-$ state due to its overlap with the $I=0^-$ state.}
%     \label{fig:LevelScheme}
% \end{figure*}

% \begin{figure}[htb]
%     \centering
%     \includegraphics[width = \columnwidth]{Q_odd-evenAg-In.png}
%     \caption{Spectroscopic quadrupole moments for even-A silver (red) \cite{Degroote2024} and indium (blue) \cite{Vernon2022,Karthein2024}. Only the even-A isomers are shown as these have the same spin, removing the spin dependence. \textbf{The errorbars on N = 66-74 are quite big as the ones quoted in the table on Ruben's PLB are both systematic and statistical added. If it is preferred I could use my own values which i know separately, or ask ruben for his separated values}}
%     \label{fig:Q_oddevenAgIn}
% \end{figure}

The following section discusses the charge radii and spectroscopic quadrupole moments of the different silver isotopes measured in this work. The charge radii are compared to neighboring chains and the quadrupole moments to empirical single-particle moments. The empirical single-particle moments are calculated with the addition rule from Ref. \cite{Heyde1994}, given by:

\begin{equation}
\begin{split}
    Q(I) = \begin{pmatrix}
        I & 2 & I\\
        -I & 0 & I
        \end{pmatrix} (-1)^{I_\pi+I_\nu+I} (2I+1) \\
        \times \left[ \begin{Bmatrix}
                    I_\pi & I & I_\nu\\
                    I & I_\pi & 2
                    \end{Bmatrix} \frac{Q(I_\pi)}{\begin{pmatrix}
        I_\pi & 2 & I_\pi\\
        -I_\pi & 0 & I_\pi
        \end{pmatrix}}\right. \\
        + \left. \begin{Bmatrix}
                    I_\nu & I & I_\pi\\
                    I & I_\nu & 2
                    \end{Bmatrix} \frac{Q(I_\nu)}{\begin{pmatrix}
        I_\nu & 2 & I_\nu\\
        -I_\nu & 0 & I_\nu
        \end{pmatrix}} \right].
\end{split}
\end{equation}

In this formula, we take the quadrupole moment of a neighboring odd-$Z$, even-$N$ isotope for $Q_\pi$, and of a neighboring even-$Z$, odd-$N$ isotope for $Q_\nu$ to calculate the empirical single-particle quadrupole moment of the odd-$Z$, odd-$N$ configuration, similar to Eq. \ref{Eq:dipole}. The $()$ and $\{\}$ indicate the Wigner 3-j and Wigner 6-j symbols respectively.

\begin{figure}[htb]
    \centering
    \includegraphics[width = \columnwidth]{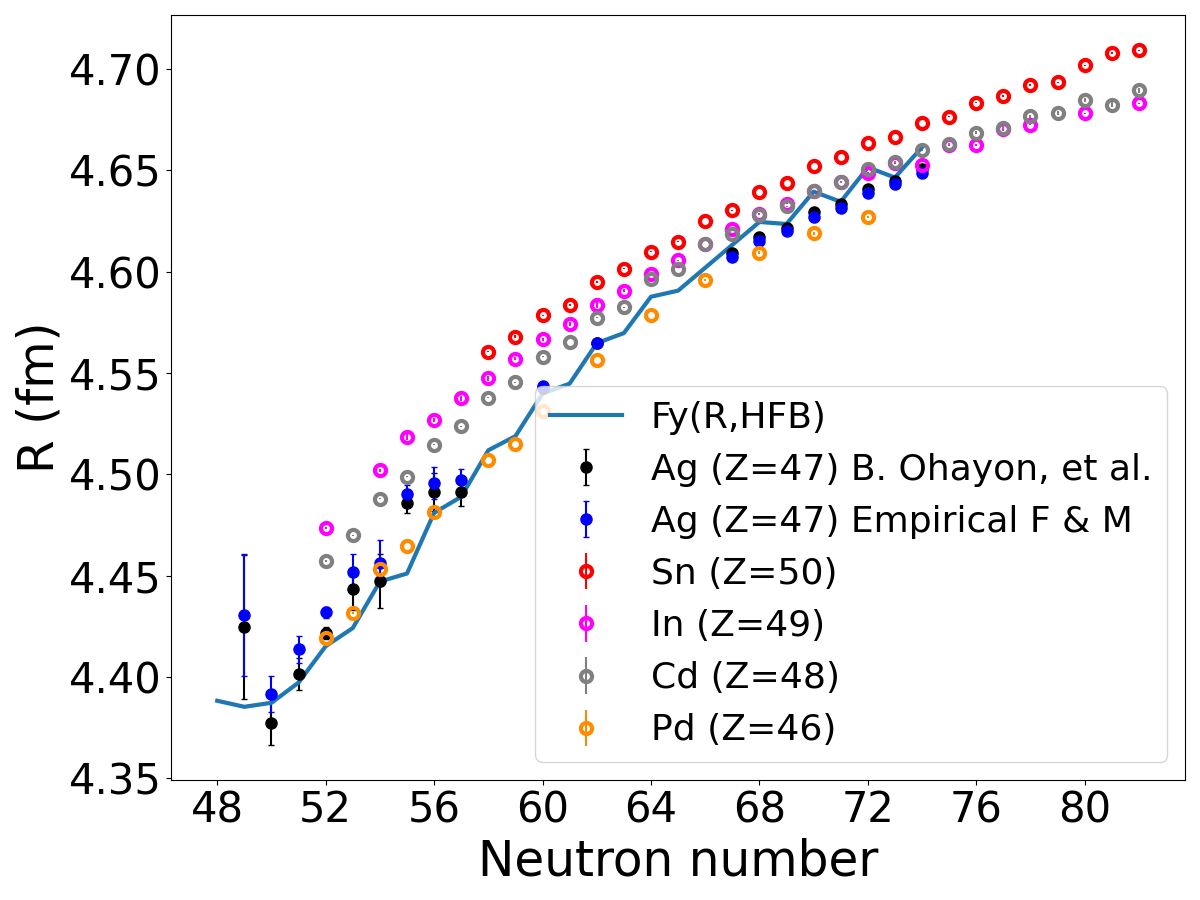}
    \caption{Charge radii for Pd, Ag, Cd, In, and Sn isotopic chains up to $N=82$. The uncertainties only denote the experimental statistical uncertainty. The blue line indicates the Fayans DFT Hartree-Fock-Bogliubov (HFB) calculations for Ag \cite{Reponen2021}, the black circles are Ag charge radii using F and M factors from Ref. \cite{Ohayon2024}, the blue circles Ag charge radii using F and M factors from Ref. \cite{Reponen2021}, the red circles Sn charge radii from Refs. \cite{Anselment1986,Piller1990}, the pink circles In charge radii from Refs. \cite{Eberz1987,Karthein2024}, the grey circles Cd charge radii from Ref. \cite{Hammen2018}, and the orange circles Pd charge radii from Ref. \cite{Geldhof2022}.}
    \label{fig:chargeradiiAll}
\end{figure}

Fig. \ref{fig:chargeradiiAll} shows the absolute charge radii of Pd ($Z=46$) up to Sn ($Z=50$) with two different sets of F and M factors for the Ag isotopic chain. The empirical F and M factors \cite{Reponen2021} are $F=-4300(300)$\,MHz/fm$^2$ and $M=1956(360)$\,GHz\,u, and the analytical-response relativistic coupled-cluster (AR-RCC) F and M factors from \cite{Ohayon2024} are $F=-3557(49)$\,MHz/fm$^2$ and $M=1479(14)$\,GHz\,u. There is a 3\,$\sigma$ difference in the field shift and a 2\,$\sigma$ difference in mass shift factors between the empirical and ab-initio calculation. In Fig. \ref{fig:chargeradiiAll}, only a small difference can be seen between the two trends on the neutron-rich side. Note that in Ref. \cite{Reponen2021} the charge radii of the lowest-spin states were reported, while here in Fig. \ref{fig:chargeradiiAll} the charge radii for the ground states are shown.
% However, on the neutron-deficient side, the difference is larger. The charge-radii trend obtained using the ab-initio results aligns better with the Pd chain.

\begin{figure}[htb]
    \centering
    \includegraphics[width = \columnwidth]{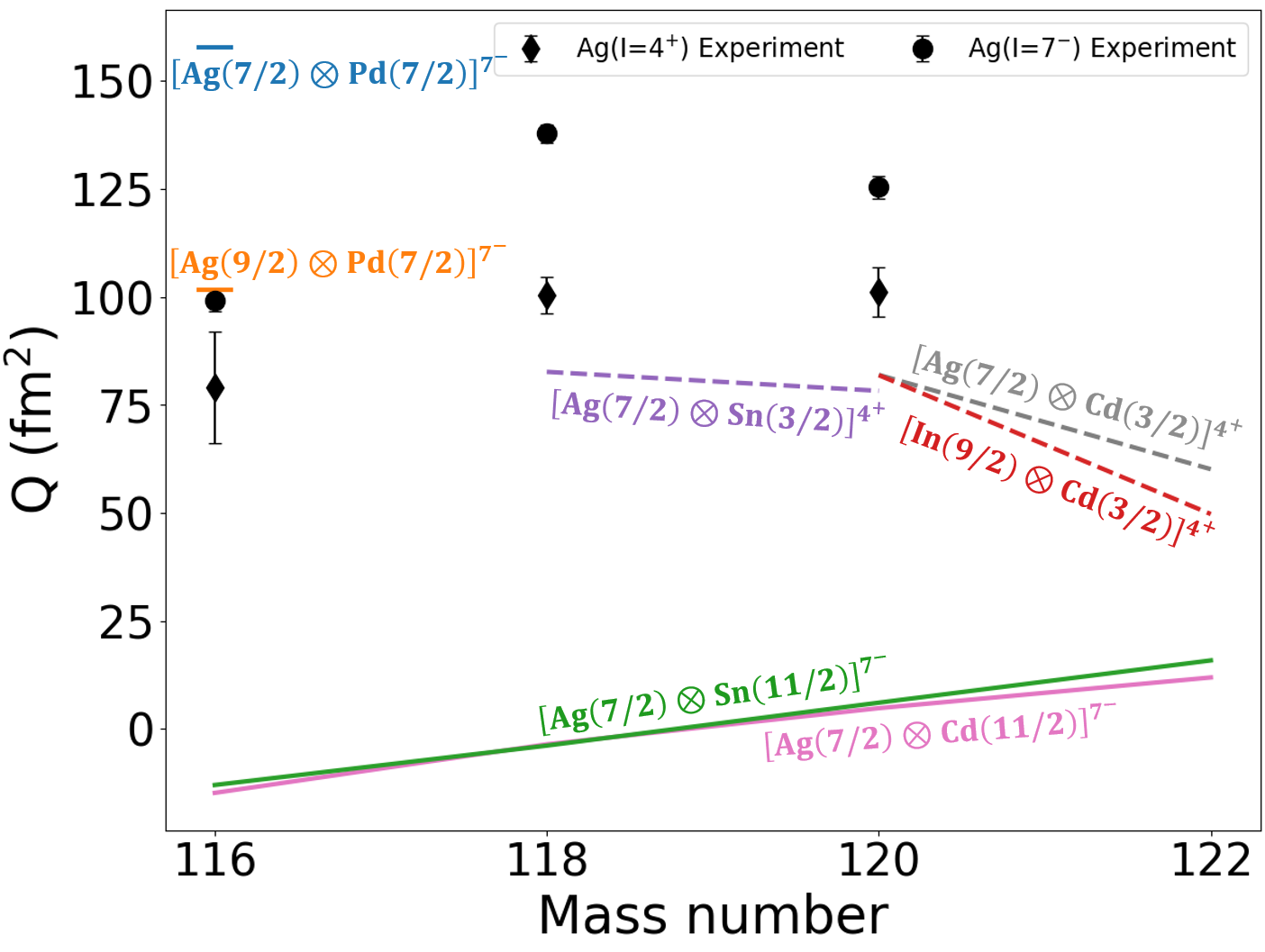}
    \caption{Spectroscopic quadrupole moments for odd-odd $I=4^+$ (black diamonds) and $I=7^-$ (black dots) states in silver with possible empirical single-particle moments. The quadrupole moments for Pd(I=7/2) are taken from Ref. \cite{Ortiz2023}, for Ag(I=7/2) from Ref. \cite{Degroote2024}, for Cd(I=3/2,11/2) from Ref. \cite{Yordanov2013}, for In(I=9/2) from Ref. \cite{Eberz1987}, and for Sn(3/2,11/2) from Refs. \cite{Dimmling1975,Anselment1986,Leblanc2004}. The error bars only contain the statistical experimental errors.}
    \label{fig:Q_oddoddAg}
\end{figure}

The spectroscopic quadrupole moments for odd-odd silver are shown in Fig. \ref{fig:Q_oddoddAg}. The $I=4^+$ and $I=7^-$ states show a positive spectroscopic quadrupole moment similar to the indium \cite{Vernon2022,Karthein2024} and odd-A silver \cite{Degroote2024} isotopes. The $I=7^-$ states do not agree very well with the empirical quadrupole moments calculated with the unpaired neutron in Cd or Sn, which show a linear increase with increasing $N$ as in odd-A Cd isotopes \cite{Yordanov2013}. However, there is an excellent agreement with the empirical moment for $\left[\text{Ag(9/2)}\otimes \text{Pd(7/2)}\right]^{7-}$, interpreted as the $[$Ag($\pi$g$_{9/2}) \otimes$Pd$(I=7/2)]^{7^-}$ configuration. This is in contrast with the trend for the g-factors in Fig. \ref{fig:gfac_spin7} which shows a relatively good agreement for both configurations. 

The empirically calculated quadrupole moments for the $I=4^+$ states agree relatively well with the experimentally measured moments. However, there seems to be no preference for a single- or three-hole proton configuration in the g$_{9/2}$ orbital, similar to the $I=7^-$ state. A more detailed interpretation of these moments would require e.g. large-scale shell model calculations.

\section{Conclusion}

The odd-odd isotopes $^{114-120}$Ag have been studied with collinear laser spectroscopy and PI-ICR mass spectrometry at IGISOL in Jyv\"askyl\"a, Finland. All spins, parities, electromagnetic moments, and level ordering of the long-lived states for $^{114-120}$Ag are determined unambiguously. Specifically, the spins of $^{116}$Ag have been reassigned. Furthermore, a new isomer has been identified in $^{118}$Ag, whose presence reinterprets the level ordering, half-lives, and spins previously suggested in the literature. The existing decay scheme of $^{118}$Ag has been shown to be incorrect. We thus identify a need to perform decay spectroscopy of isomerically purified beams in a future experiment. Lastly, the previously unknown excitation energy of the low-spin isomer in $^{120}$Ag has been measured. 

The charge radii were found to vary smoothly with N, similar to the neighboring elements. From the magnetic dipole and quadrupole moments, it can be concluded that these isotopes have a rather mixed neutron configuration, in line with the more collective properties of Pd. This illustrates that silver sits at the transition between the spherical Sn isotopes, and the well-deformed isotopic chains $Z<47$.

Multiple campaigns are ongoing to understand this region further. A decay spectroscopy experiment has been performed at IGISOL, Jyv\"askyl\"a to reassign the $\gamma$-transitions in the decay of $^{118}$Pd. An experimental campaign at CRIS at ISOLDE/CERN is ongoing to perform laser spectroscopy on neutron-rich silver, approaching the $N=82$ magic shell closure \cite{IS660}. Moreover, a muonic X-ray experiment on $^{107,108m,109}$Ag at PSI is being prepared to establish high-precision field and mass shift factors \cite{Cocolios2020}. 

More work is also being done from the theoretical perspective. Atomic \textit{ab initio} calculations are ongoing to calculate the hyperfine $A$ and $B$ constants in the 5s\,$^2S_{1/2}$, 5p\,$^2P_{1/2}$ and 5p\,$^2P_{3/2}$ atomic levels and the field and mass shift factors in the 4d$^{10}$5s\,$^2S_{1/2}\rightarrow$\,4d$^{10}$5p\,$^2P_{1/2}$ and 4d$^{10}$5s\,$^2S_{1/2}\rightarrow$\,4d$^{10}$5p\,$^2P_{3/2}$ transitions. This is calculated in parallel with new state-of-the-art DFT calculations for the diamagnetic and paramagnetic shielding constants to revise the reference magnetic dipole moment from Ref. \cite{Sahm1974}.

This work provides a step towards understanding the nuclear structure in the largely unexplored region moving away from the spherical Sn isotopes. 

\begin{acknowledgments}

% \textbf{AK: The text below was copied from the odd-A silver mass+lasers paper from IGISOL but I commented out Jacek's grants.}
%This work was partially supported by the STFC Grant Nos.\,ST/M006433/1, \,ST/P003885/1, ST/V001035/1, ST/P004598/1 and ST/P004423/1, and by the Polish National Science Centre under Contract No.\,2018/31/B/ST2/02220. 

%We acknowledge the CSC-IT Center for Science Ltd., Finland, for the allocation of computational resources. This project was partly undertaken on the Viking Cluster, which is a high performance compute facility provided by the University of York. We are grateful for computational support from the University of York High Performance Computing service, Viking and the Research Computing team.

RPDG received funding from the European Union’s Horizon 2020 research and innovation programme under the Marie Sk{\l}odowska-Curie grant agreement No 844829. The funding from the European Union’s Horizon 2020 research and innovation program under grant agreement No. 771036 (ERC CoG MAIDEN) and Research Council of Finland (Grant Nos. 314733, 320062, 318043, 295207, 327629 and 354968) are gratefully acknowledged. This work was supported by the UK Science and Technology Facilities Council (STFC). We acknowledge W. N\"ortersh\"auser for the use of the charge-exchange cell. 

\end{acknowledgments}

\bibliography{bibfile}

\newpage

\title{Supplementary Material: Binding energies, charge radii, spins and moments: odd-odd Ag isotopes and discovery of a new isomer}

\maketitle

\section{Additional results on odd-even Ag}\label{SupMatSec:odd-even}

Results obtained from the same experimental campaign as this work on odd-even $^{113-121}$Ag have been partially published before in Refs. \cite{Reponen2021,Degroote2024}. Here we provide the $A$-parameters, isotope shifts, and changes in mean-squared radii not published before (see Table \ref{SUPtab:laserresults}).

\begin{table}[htb]
    \caption{Complementary results of the laser spectroscopy study from Ref. \cite{Reponen2021,Degroote2024} with their spins and parities $I^{\pi}$. Columns $A$ and $\delta\nu$ show the hyperfine $A$-constants and isotope shifts of the hyperfine spectra. $\delta\langle$r$^2\rangle^{109,A}$ shows the change in mean-squared radii, calculated using $^{109}$Ag as a reference. Statistical and systematic errors are given in normal and square brackets, respectively.}
    \label{SUPtab:laserresults}
    \begin{threeparttable}
    \begin{tabular}{lllll}
        \toprule
Nuclide & $I^{\pi}$ & A(P$_{3/2}$)  & $\delta\nu$  & $\delta\langle$r$^2\rangle^{109,A}$  \\
 &  & (MHz) &  (MHz) & (fm$^2$) \\
\hline
$^{113}$Ag$^m$ 	    &$7/2^+$ & 173.7(13) & $-650(5)$   &  0.3147(14)[45]           	\\
$^{115}$Ag$^m$ 	    &$7/2^+$ & 173.5(7)  & $-888(8)$   &  0.446(2)[6]           	\\
$^{117}$Ag$^m$ 	    &$7/2^+$ & 172.5(4)  & $-1077(18)$ &  0.561(5)[8]           	\\
$^{119}$Ag$^m$ 	    &$7/2^+$ & 177.2(5)  & $-1249(26)$ &  0.669(7)[10]           	\\
$^{121}$Ag 	        &$7/2^+$ & 172.5(16) & $-1355(5)$  &  0.7570(14)[110]           	\\
        \hline
    \end{tabular}
\end{threeparttable}
\end{table}

\section{Extraction of relative production yields from hyperfine spectra}\label{SupMatSec:production}

The relative production yields were approximated by comparing the integration of hyperfine peaks, i.e. transitions, for different nuclear states. First, the background was subtracted from every spectrum. Then, $M_j$ which is the sum of the integrals of all hyperfine transitions for all hyperfine spectra of a nuclear state $j$, is calculated with Eq. \ref{Eq:Mj}. $M_j$ is corrected for the acquisition time and for Racah intensity via the NRacah$_h$ factor defined in Eq. \ref{Eq:NRacah}. The idea behind Eq. \ref{Eq:Mj} is that nuclear states with the same production rate will result in the same $M_j$ insensitive to the nuclear spin \textit{I}, the chosen transition \textit{h}, and the acquisition time \textit{i}. The relative yield was calculated using Eq. \ref{Eq:relyield}.

\begin{eqnarray}
    \label{Eq:Mj}
    M_j = \sum_{h=0}^{m=\#\text{transitions}}\frac{1}{\text{NRacah}_h}\times\sum_{i=0}^{N_s}\frac{\text{integral}_{hi}}{\text{time}_i} \\
    \label{Eq:NRacah}
    \text{NRacah}_h = \frac{\text{Racah}_h}{\sum_{k=\text{all transitions}}\text{Racah}_k}\\
    \label{Eq:relyield}
    \text{Relative yield}_I = \frac{M_I}{\sum_{j=0,4,7} M_j}\,\,\, 
\end{eqnarray}
\\
The integral$_{hi}$ is the integral of the transition \textit{h} in the spectrum \textit{i}, $N_s$ is the total number of scans of a nuclear state $j$, NRacah$_h$ is the normalized Racah intensity of the transition \textit{h} and time$_i$ is the acquisition time of the spectrum \textit{i}. Then to retrieve the relative yield of a nuclear state \textit{I} Eq. \ref{Eq:relyield} was used, where the sum is over the spins of all measured nuclear states. In total, there were four hyperfine spectra measured and eight different sets of transitions were evaluated. The statistical error on each set is derived by calculating the standard deviation between $\frac{\text{integral}_{hi}}{\text{time}_i}$ for the different spectra \textit{i}.

These results are shown in Table \ref{SUPtab:production} together with the ratio for the assigned level order. We note there is some spread on the obtained results in Table \ref{SUPtab:production}, likely due to the Racah intensity approximation used in the calculation.

\begin{table*}[htb]
    \caption{Results of the relative production yields with the first column showing the chosen transitions for each state, the second column showing the relative production yields, and the last column the ratio of the assigned level order.}
    \label{SUPtab:production}
    \begin{threeparttable}
    \begin{tabular}{|lc|ccc|c|}
        \toprule
\multicolumn{2}{|c|}{Chosen hyperfine transitions to integrate} & \multicolumn{3}{|c|}{Relative production yield (\%)} & Ratio\\
\multicolumn{1}{|c}{$I=4^+$} & $I=7^-$  & $I=0^-$ & $I=4^+$ & $I=7^-$  & Spin $4/(1+7)$ \\
\hline
$9/2\rightarrow11/2 + 7/2\rightarrow5/2$ & $15/2\rightarrow17/2 + 13/2\rightarrow11/2$ & 8.9(10) & 30.8(35) & 60(4) &  0.45(6)  \\
$9/2\rightarrow11/2 + 7/2\rightarrow5/2$ & $15/2\rightarrow15/2 + 13/2\rightarrow11/2$ & 10.6(10) & 36.5(35) & 53(4) &  0.57(7)   \\
$9/2\rightarrow11/2 + 7/2\rightarrow5/2$ & $15/2\rightarrow17/2 + 13/2\rightarrow13/2$ & 9.3(10) & 32.1(35) & 59(4) &  0.47(6)  \\
$9/2\rightarrow11/2 + 7/2\rightarrow5/2$ & $15/2\rightarrow15/2 + 13/2\rightarrow13/2$ & 11.1(10) & 38.3(33) & 50.6(33) &  0.62(6)  \\
$9/2\rightarrow9/2 + 7/2\rightarrow5/2$ & $15/2\rightarrow17/2 + 13/2\rightarrow11/2$ & 9.5(11) & 26.4(29) & 64.0(35) &  0.36(4)  \\
$9/2\rightarrow11/2 + 7/2\rightarrow7/2$ & $15/2\rightarrow17/2 + 13/2\rightarrow11/2$ & 9.3(10) & 28.1(34) & 63(4) &  0.39(5)   \\
$9/2\rightarrow9/2 + 7/2\rightarrow7/2$ & $15/2\rightarrow17/2 + 13/2\rightarrow11/2$ & 9.9(11) & 23.5(26) & 66.6(34) &  0.31(4)   \\
$9/2\rightarrow9/2 + 7/2\rightarrow7/2$ & $15/2\rightarrow15/2 + 13/2\rightarrow13/2$ & 12.6(11) & 30.0(24) & 57.4(29) &  0.42(4)   \\
        \hline
\multicolumn{2}{|c|}{\textbf{Average}} & \textbf{10.2(4)} & \textbf{30.7(11)} & \textbf{59.1(13)} & \textbf{0.443(18)}  \\
\hline
    \end{tabular}
\end{threeparttable}
\end{table*}

\end{document}